\documentclass[twocolumn,journal]{IEEEtran}
\usepackage[T1]{fontenc}
\usepackage[latin9]{inputenc}
\usepackage{array}
\usepackage{verbatim}
\usepackage{prettyref}
\usepackage{float}
\usepackage{multirow}
\usepackage{amsmath}
\usepackage{amsthm}
\usepackage{amssymb}
\usepackage{graphicx}
\usepackage[unicode=true,
 bookmarks=true,bookmarksnumbered=true,bookmarksopen=true,bookmarksopenlevel=1,
 breaklinks=false,pdfborder={0 0 0},pdfborderstyle={},backref=false,colorlinks=false]
 {hyperref}
\hypersetup{pdftitle={Your Title},
 pdfauthor={Your Name},
 pdfpagelayout=OneColumn, pdfnewwindow=true, pdfstartview=XYZ, plainpages=false}

\makeatletter

\providecommand{\tabularnewline}{\\}
\floatstyle{ruled}
\newfloat{algorithm}{tbp}{loa}
\providecommand{\algorithmname}{Algorithm}
\floatname{algorithm}{\protect\algorithmname}

\theoremstyle{plain}
\newtheorem{thm}{\protect\theoremname}
\theoremstyle{definition}
\newtheorem{problem}[thm]{\protect\problemname}
\theoremstyle{remark}
\newtheorem{rem}[thm]{\protect\remarkname}

\usepackage[caption=false,font=footnotesize]{subfig}
\usepackage{cite}

\@ifundefined{showcaptionsetup}{}{%
 \PassOptionsToPackage{caption=false}{subfig}}
\usepackage{subfig}
\makeatother

\providecommand{\problemname}{Problem}
\providecommand{\remarkname}{Remark}
\providecommand{\theoremname}{Theorem}

\begin{document}
\title{Maximum A Posteriori Probability (MAP) Joint Fine Frequency Offset
and Channel Estimation for MIMO Systems with Channels of Arbitrary
Correlation}
\author{Mingda Zhou, Zhe Feng, Xinming Huang, Youjian (Eugene) Liu\thanks{Zhe Feng and Youjian (Eugene) Liu are with the Department of Electrical,
Computer, \& Energy Engineering, University of Colorado at Boulder,
e-mail: \protect\href{mailto:eugeneliuieee\%40ieee.org}{eugeneliuieee@ieee.org}.
The work was partially supported by US NSF grant ECCS-1408604 and
IIP-1414250.}\thanks{Mingda Zhou and Xinming Huang are with the Department of Electrical
and Computer Engineering, Worcester Polytechnic Institute, e-mail:
\protect\href{mailto:xhuang\%40wpi.edu}{xhuang@wpi.edu}.}}
\maketitle
\begin{abstract}
Channel and frequency offset estimation is a classic topic with a
large body of prior work using mainly maximum likelihood (ML) approach
together with Cramér-Rao Lower bounds (CRLB) analysis. We provide
the maximum a posteriori (MAP) estimation solution which is particularly
useful for for tracking where previous estimation can be used as prior
knowledge. Unlike the ML cases, the corresponding Bayesian Cramér-Rao
Lower bound (BCRLB) shows clear relation with parameters and a low
complexity algorithm achieves the BCRLB in almost all SNR range. We
allow the time invariant channel within a packet to have arbitrary
correlation and mean. The estimation is based on pilot/training signals.
An unexpected result is that the joint MAP estimation is equivalent
to an individual MAP estimation of the frequency offset first, again
different from the ML results. We provide insight on the pilot/training
signal design based on the BCRLB. Unlike past algorithms that trade
performance and/or complexity for the accommodation of time varying
channels, the MAP solution provides a different route for dealing
with time variation. Within a short enough (segment of) packet where
the channel and CFO are approximately time invariant, the low complexity
algorithm can be employed. Similar to belief propagation, the estimation
of the previous (segment of) packet can serve as the prior knowledge
for the next (segment of) packet.
\end{abstract}


\section{Introduction\label{sec:Introduction}}

We consider joint carrier frequency offset (CFO) and channel coefficient
estimation for multiple-input-multiple-output (MIMO) flat fading channels.
In addition to being a critical part of a communication system, the
solution has applications in other systems. For example, in radar
systems, the CFO is related to Doppler frequency and can be used to
estimate target speed and the channel coefficient estimation of an
antenna array can be used to estimate target direction.

This is a classic problem with a large body of prior workusing mainly
maximum likelihood (ML) estimation approach together with Cramér-Rao
Lower bounds (CRLB) analysis. The maximum a posteriori (MAP) estimation
solution, low complexity algorithms, and the corresponding Bayesian
Cramér-Rao lower bound (BCRLB) for this problem has not appeared in
literature. We provide the result here so that future designers can
choose between the MAP and ML approaches depending on the trade-offs
in a system, especially for tracking that uses previous estimation
as prior knowledge.

\subsection{Contributions}

In this work, we allow the channel to have arbitrary spatial correlation
and mean. After subtracting the mean, it has  a circularly symmetric
complex Gaussian distribution. While the channel is assumed to be
time invariant for the estimation problem, the MAP result provides
a different approach to deal with time varying channels than past
literature. It is assumed that the coarse frequency synchronization
has been done so that the discrete time model for the matched filter
output is valid for a fine frequency offset. The estimation is based
on pilot/training signals. The simple model leads to clean results
and low complexity algorithm that achieve the BCRLB in almost all
SNR range. The contributions of the paper are listed below.
\begin{enumerate}
\item We provide the solution for the joint MAP frequency offset and channel
estimation. An \emph{unexpected} result is that the joint MAP estimation
is equivalent to an individual MAP estimation of the frequency offset
first with only the channel statistical information, followed by an
MMSE estimation of the channel with the estimated frequency offset
substituted in. This is different from the past joint maximum likelihood
(ML) estimation results, where the joint estimation is not equivalent
to individual estimation. In addition, the MAP solution includes the
ML solution as a special case when we let the variances of the CFO
and channel approach infinity.
\item The Bayesian Cramér-Rao Lower bound (BCRLB) is derived in closed form
for the frequency offset estimation with prior knowledge. Unlike the
complicated CRLB bound for joint ML CFO and channel estimation \cite{Stoica_2003ITSP_ParameterEstimationMIMOFlatfadingChannelsFrequencyOffsets},
the BCRLB exhibits explicit and easy-to-understand relation to various
parameters and does not depend on the channel realization. %
\item %
{} Therefore, the BCRLB provides new insight on the pilot/training signal
design, including the effect of time spreading, and structures of
periodic pilot and time division pilot.
\item A closed form low complexity high performance algorithm that does
not need search is provided. Numerical results has demonstrated that
the algorithm achieves the BCRLB in almost all SNR range, while past
ML algorithms perform poorly in low SNR range. The algorithm is demonstrated
to achieve maximum acquisition range allowed by the discrete time
model and the pilot structure
\item Unlike past algorithms that trade performance and/or complexity for
the accommodation of time varying channels, we provide a different
route for dealing with time variation. Within a short enough (segment
of) packet where the channel and CFO are approximately invariant,
the low complexity algorithm can be employed. Similar to belief propagation,
the estimation of the previous (segment of) packet can serve as the
prior knowledge for the next (segment of) packet. %
\end{enumerate}

\subsection{Related Work}

Frequency estimation is a classic problem. For single-input-single-output
(SISO) systems in additive white Gaussian noise (AWGN) channels, an
early paper on ML estimation of frequency, phase, and amplitude of
a single tone from discrete time samples of the output of an AWGN
channel is \cite{Boorstyn_1974ITIT_SingleToneParameterEstimationDiscretetimeObservations},
where search algorithms taking advantage of FFT and the CRLB is provided.
Another ML estimator for AWGN channel is proposed in \cite{Reggiannini_1995ITC_CarrierFrequencyRecoveryAlldigitalModemsBurstmodeTransmissions},
where a suboptimal algorithm that only uses the phases of the estimated
autocorrelation function of the received signal is given. The algorithm
is applied to a satellite communication system and a GSM communication
system, whose models are both made close to the AWGN channel.

The frequency offset estimation for SISO flat fading channel has been
well investigated. In \cite{Fitz_1997IToC_FrequencyOffsetCompensationPilotSymbolAssistedModulationFrequencyFlatFading},
the maximum-likelihood (ML) estimator of frequency offset is given
for pilot aided communications in a time varying fading channel. The
approximation $\sin(z)\doteq z$ is used to approximately solve for
a stationary point of the ML metric. It only utilizes a small lag
to avoid phase unwrapping, which leads to a degradation of the performance.
Newton search and local grid search were also applied to refine the
estimate, where the Newton search does not work well because of local
maximums, and the accuracy of grid search depends heavily on resolution
and search range. A low complexity suboptimal algorithm that only
employs the phase of the autocorrelation of the matched filter is
also proposed, which is modified in \cite{Vitetta_1998ICL_FurtherResultsCarrierFrequencyEstimationTransmissionsFlatFadingChannels},
where the difference of adjacent phases is used to replace the phases
to avoid phase unwrapping and to increase the acquisition range. %
{} %
{} The algorithms of \cite{Fitz_1997IToC_FrequencyOffsetCompensationPilotSymbolAssistedModulationFrequencyFlatFading},
\cite{Vitetta_1998ICL_FurtherResultsCarrierFrequencyEstimationTransmissionsFlatFadingChannels}
are further modified in \cite{Stoica_2001ICL_FrequencyOffsetEstimationFlatfadingChannels}
to improve the modeling error of the time varying fading process.
The first method uses equal weighting to avoid dependence on the fading
process. The second method estimates the frequency offset and the
fading correlation jointly, resulting in low complexity of a square
operation and a grid search of the output of an FFT.

For time invariant MIMO flat fading, the ML joint estimation of the
channel and frequency offsets has been studied in a comprehensive
work \cite{Stoica_2003ITSP_ParameterEstimationMIMOFlatfadingChannelsFrequencyOffsets}.
The frequency offsets between pairs of transmit and receive antennas
are allowed to be different. It is shown that the CRLB for the channel
and CFO estimation depends on the true value of the channel and CFOs
in a complicated manner. Simplified bounds for stationary pilot in
the limit of infinite long transmission is provided. In general, optimal
pilot signals depend on the channel. The estimation algorithm for
the general case is a $n$-dimensional search where $n$ is the number
of CFOs. For specially designed orthogonal pilot signal, where one
antenna is active at one symbol time, the $n$-dimensional search
can be converted to $n$ 1-D problems. Both pilot signal based (data
aided) and decision statistics feedback (code-aided) based joint single
frequency offset and channel estimations by ML are considered in \cite{Moeneclaey_2006ITWC_ReducedComplexityDataaidedCodeaidedFrequencyOffsetEstimationFlatfadingMIMOChannels}.
The pilot based case is similar to that of \cite{Stoica_2003ITSP_ParameterEstimationMIMOFlatfadingChannelsFrequencyOffsets}
when specialized to a single frequency case, where orthogonal pilot
with orthogonal rows and columns is used to achieve zero self-noise.
The work recognizes the benefit of orthogonal periodic pilot signals.
Our algorithm includes the algorithm in this paper as a special case.
The code-aided case employs expectation maximization (EM) algorithm.
Iterative EM is also employed in \cite{Liang_2008IToWC_JointChannelFrequencyOffsetEstimationDistributedMIMOFlatfadingChannels},
where the same setting as in \cite{Stoica_2003ITSP_ParameterEstimationMIMOFlatfadingChannelsFrequencyOffsets}
is considered, in order to avoid the pilot structure in \cite{Stoica_2003ITSP_ParameterEstimationMIMOFlatfadingChannelsFrequencyOffsets}
where one antenna is active at one symbol time. The performance is
close to the CRLB derived in \cite{Stoica_2003ITSP_ParameterEstimationMIMOFlatfadingChannelsFrequencyOffsets}.
The only work related to MAP estimation that we found is \cite{Tarokh_2008ITSP_BoundsAlgorithmsFrequencySynchronizationCollaborativeCommunicationSystems}
for relay networks where Bayesian Cramér-Rao Lower Bound is used and
the frequency is assumed to be Gaussian distributed. CFO estimation
for other settings has been studied, such as MIMO frequency selective
fading channels with OFDM modulation \cite{Stuber_2001IGTC2G'_SynchronizationMIMOOFDMSystems,Ng_2004ITSP_SemiblindChannelEstimationMethodMultiuserMultiantennaOFDMSystems,Wang_2005ITC_EMbasedIterativeReceiverDesignCarrierfrequencyOffsetEstimationMIMOOFDMSystems,Kuo_2006IToC_MaximumlikelihoodSynchronizationChannelEstimationOFDMAUplinkTransmissions,Li_2006ITC_OptimalTrainingSignalsMIMOOFDMChannelEstimationPresenceFrequencyOffsetPhaseNoise,Ng_2007ITC_JointSemiblindFrequencyOffsetChannelEstimationMultiuserMIMOOFDMUplink,Ko_2009IToVT_JointChannelEstimationSynchronizationMIMOOFDMPresenceCarrierSamplingFrequencyOffsets,Berbineau_2011IToVT_JointCarrierFrequencyOffsetFastTimeVaryingChannelEstimationMIMOOFDMSystems,Hari_2012CN2NCO_JointEstimationSynchronizationImpairmentsMIMOOFDMSystem,Moretti_2013IWCL_JointMaximumLikelihoodEstimationCFONoisePowerSNROFDMSystems,Orozco-Lugo_2013IWCL_CarrierFrequencyOffsetEstimationOFDMAUsingDigitalFiltering,Gao_2016ITWC_ComputationallyEfficientBlindEstimationCarrierFrequencyOffsetMIMOOFDMSystems},
multi-user \cite{Kennedy_2011IC_BlindTimingCarrierSynchronisationDistributedMultipleInputMultipleOutputCommunicationSystems,Oteri_2015CNCI2ICO_CarrierFrequencyOffsetCorrectionUplinkMultiuserMIMONextGenerationWiFi},
and multi-hop networks \cite{Blostein_2011ITWC_BoundsAlgorithmsMultipleFrequencyOffsetEstimationCooperativeNetworks}.

The rest of this paper is organized as follows. Section \ref{sec:System-Model}
provides the system model. In Section \ref{sec:The-Optimization-Problem},
the joint MAP estimation of CFO and channel is shown to be separable.
In Section \ref{sec:Fine-Frequency-Synchronization}, the frequency
synchronization algorithm is designed. To analyze the performance
limit, BCRLBs as design guidelines are derived in Section \ref{sec:Performance-Analysis},
where the pilot signal design is discussed. In Section \ref{sec:Simulation-Results},
we show simulation results of the proposed algorithm in terms of estimation
error variance and acquisition range. Results for time varying channel
is also given. Section \ref{sec:Conclusion} concludes.

\emph{Notation Convention:} We use our notation convention in Table
\ref{tab:Notations}. It is convenient for organizing variables with
multiple indices into matrices or vectors or vectorizing a matrix.

\begin{table}[h]
\caption{\label{tab:Notations}Notation Convention}

\centering{}%
\begin{tabular*}{0.95\columnwidth}{@{\extracolsep{\fill}}|>{\raggedright}m{0.2\columnwidth}|>{\raggedright}p{0.65\columnwidth}|}
\hline 
Notation & Meaning\tabularnewline
\hline 
\hline 
$x$ & a scalar\tabularnewline
\hline 
$\vec{x}$ & a \emph{column} vector\tabularnewline
\hline 
$X$ & a matrix\tabularnewline
\hline 
$\boldsymbol{x}$, $\vec{\boldsymbol{x}}$, $\boldsymbol{X}$ & a random variable, column random vector, random matrix\tabularnewline
\hline 
\multirow{2}{0.2\columnwidth}{$[a_{x_{1},x_{2}}]_{x_{1},x_{2}}$} & a matrix whose element at $x_{1}$-th row and $x_{2}$-th column is
$a_{x_{1},x_{2}}$, e.g., $\left[\begin{array}{cc}
a_{1,1} & a_{1,2}\\
a_{2,1} & a_{2,2}
\end{array}\right]=\left[a_{i,j}\right]_{i,j}$ ; \tabularnewline
 & $x_{1}$ or $x_{2}$ can be continuous variables\tabularnewline
\hline 
\multirow{2}{0.2\columnwidth}{$[a_{x}]_{x,x}$} & a diagonal matrix whose element at $x$-th row and $x$-th column
is $a_{x}$\tabularnewline
 & other elements are zero, e.g., $\left[\begin{array}{cc}
a_{1} & 0\\
0 & a_{2}
\end{array}\right]=\left[a_{i}\right]_{i,i}$\tabularnewline
\hline 
$\left[A_{x_{1},x_{2}}\right]_{x_{1},x_{2}}$  & a block matrix whose block at $x_{1}$-th row and $x_{2}$-th column
is $A_{x_{1},x_{2}}$\tabularnewline
\hline 
$\left[\vec{a}_{x}\right]_{1,x}$ & a matrix whose $x$-th column is $\vec{a}_{x}$, e.g., $\left[\begin{array}{cc}
\vec{a}_{1} & \vec{a}_{2}\end{array}\right]=\left[\vec{a}_{i}\right]_{1,i}$\tabularnewline
\hline 
$[a_{x}]_{x}$ & a column vector whose element at the $x$-th row is $a_{x}$, e.g.,
$\left[\begin{array}{c}
a_{1}\\
a_{2}\\
a_{3}
\end{array}\right]=\left[a_{i}\right]_{i}$\tabularnewline
\hline 
$\left[\vec{a}_{x}\right]_{x}$ & a tall vector whose $x$-th row of vector is $\vec{a}_{x}$, e.g.,
$\left[\begin{array}{c}
a_{1,1}\\
a_{2,1}\\
a_{1,2}\\
a_{2,2}
\end{array}\right]=$ $\left[\begin{array}{c}
\left[a_{i,1}\right]_{i}\\
\left[a_{i,2}\right]_{i}
\end{array}\right]=\left[\left[a_{i,j}\right]_{i}\right]_{j}$\tabularnewline
\hline 
$\left[\vec{a}_{x}^{T}\right]_{x}$  & a matrix whose $x$-th row is $\vec{a}_{x}^{T}$, e.g., $\left[\begin{array}{c}
\vec{a}_{1}^{T}\\
\vec{a}_{2}^{T}
\end{array}\right]=\left[\vec{a}_{i}^{T}\right]_{i}$\tabularnewline
\hline 
$\tilde{x}(f)$ & Fourier transform of $x(t)$, i.e., $\tilde{x}(f)=\mathcal{F}\left\{ x(t)\right\} (f)$.\tabularnewline
\hline 
\end{tabular*}
\end{table}

\section{System Model\label{sec:System-Model}}

We investigate time invariant joint CFO and flat fading channel estimation
for MIMO systems. The transmitter has $l_{\text{t}}$ antennas and
the receiver has $l_{\text{r}}$ antennas. The received signal of
the $r$-th receive antenna at the $k$-th symbol time is modeled
as 
\begin{eqnarray*}
\boldsymbol{y}_{r,k} & = & e^{j2\pi\boldsymbol{f}_{\delta}(k-1)}\sum_{t=1}^{l_{\text{t}}}s_{t,k}\boldsymbol{h}_{r,t}+\boldsymbol{n}_{r,k},
\end{eqnarray*}
where $r=1,...,l_{\text{r}}$; $k=1,...,n$ is the symbol time index;
$\boldsymbol{h}_{r,t}\in\mathbb{C}$ is the channel coefficient from
the $t$-th transmit antenna to the $r$-th receive antenna; $\boldsymbol{n}_{r,k}\sim\mathcal{CN}\left(0,\sigma_{\boldsymbol{n}}^{2}\right)$,
$\sigma_{\boldsymbol{n}}^{2}=1$, $\forall r,k$, are i.i.d. circularly
symmetric complex Gaussian distributed with zero mean and unit variance;
$s_{t,k}\in\mathbb{C}$ is the pilot/training signal sent from the
$t$-th transmit antenna at time $k$; $\boldsymbol{f}_{\delta}=\bar{\boldsymbol{f}}_{\delta}t_{b}$
is the residual normalized carrier frequency offset (CFO) due to what
is left from the coarse frequency synchronization; $t_{b}$ is the
symbol period; $\bar{\boldsymbol{f}}_{\delta}$ is the pre-normalized
carrier frequency offset. In this paper, CFO refers to $\boldsymbol{f}_{\delta}$.
To write the model in vector form, define $\vec{\boldsymbol{y}}_{r}=\left[\boldsymbol{y}_{r,k}\right]_{k}\in\mathbb{C}^{n\times1}$,
$\vec{\boldsymbol{y}}=\left[\vec{\boldsymbol{y}}_{r}\right]_{r},$
$\vec{\boldsymbol{h}}_{r}=\left[\boldsymbol{h}_{r,t}\right]_{t=1:l_{\text{t}}}\in\mathbb{C}^{l_{\text{t}}\times1}$,
$\vec{\boldsymbol{h}}=\left[\boldsymbol{\vec{h}}_{r}\right]_{r=1:l_{\text{r}}}\in\mathbb{C}^{l_{\text{r}}l_{\text{t}}\times1}$,
$S=\left[s_{t,k}\right]_{k,t}\in\mathbb{C}^{n\times l_{\text{t}}}$,
\begin{eqnarray*}
\boldsymbol{F} & = & F(\boldsymbol{f}_{\delta})=\left[e^{j2\pi\boldsymbol{f}_{\delta}(k-1)}\right]_{k,k=1:n}\\
 & = & \left[\begin{array}{cccc}
e^{j2\pi\boldsymbol{f}_{\delta}\cdot0} & 0 & \cdots & 0\\
0 & e^{j2\pi\boldsymbol{f}_{\delta}\cdot1} & \ddots & \vdots\\
\vdots & \ddots & \ddots & 0\\
0 & \cdots & 0 & e^{j2\pi\boldsymbol{f}_{\delta}(n-1)}
\end{array}\right],
\end{eqnarray*}
$\boldsymbol{X}=\boldsymbol{F}S$, block diagonal matrix
\begin{eqnarray*}
\grave{\boldsymbol{X}} & = & \left[\boldsymbol{X}\right]_{r,r=1:l_{\text{r}}}=\left[\begin{array}{ccc}
\boldsymbol{X} & \mathbf{0} & \mathbf{0}\\
\mathbf{0} & \ddots & \mathbf{0}\\
\mathbf{0} & \mathbf{0} & \boldsymbol{X}
\end{array}\right]=\grave{\boldsymbol{F}}\grave{S},
\end{eqnarray*}
 $\grave{\boldsymbol{F}}=\left[\boldsymbol{F}\right]_{r,r}$, $\grave{S}=\left[S\right]_{r,r}$,
$\vec{\boldsymbol{n}}_{r}=\left[\boldsymbol{n}_{r,k}\right]_{k},$
and $\vec{\boldsymbol{n}}=\left[\vec{\boldsymbol{n}}_{r}\right]_{r}$.
Then we have
\begin{eqnarray*}
\vec{\boldsymbol{y}}_{r} & = & \boldsymbol{X}\vec{\boldsymbol{h}}_{r}+\vec{\boldsymbol{n}}_{r},
\end{eqnarray*}
\begin{eqnarray}
\vec{\boldsymbol{y}} & = & \grave{\boldsymbol{X}}\vec{\boldsymbol{h}}+\vec{\boldsymbol{n}}.\label{eq:channel-model-fine-freq}
\end{eqnarray}

The spatially correlated channel state $\vec{\boldsymbol{h}}$ has
distribution $\mathcal{CN}\left(\vec{\mu}_{\vec{\boldsymbol{h}}},\Sigma_{\vec{\boldsymbol{h}}}\right)$,
where $\vec{\mu}_{\vec{\boldsymbol{h}}}=\left[\left[\mu_{\boldsymbol{h}_{r,t}}\right]_{t}\right]_{r}$
is the mean; and 
\begin{eqnarray}
\Sigma_{\vec{\boldsymbol{h}}} & = & \left[\left[c_{\boldsymbol{h}_{r_{1},t_{1}},\boldsymbol{h}_{r_{2},t_{2}}}\right]_{t_{1},t_{2}}\right]_{r_{1},r_{2}}\label{eq:cov-h}
\end{eqnarray}
 is the covariance matrix of $\vec{\boldsymbol{h}}$ and $c_{\boldsymbol{h}_{r_{1},t_{1}},\boldsymbol{h}_{r_{2},t_{2}}}$
is the covariance between $\boldsymbol{h}_{r_{1},t_{1}}$ and $\boldsymbol{h}_{r_{2},t_{2}}$.
{} %
The frequency offset $\boldsymbol{f}_{\delta}$ is approximated with
Gaussian distribution $\mathcal{N}(\mu_{\boldsymbol{f}_{\delta}},\sigma_{\boldsymbol{f}_{\delta}}^{2})$.
The variance of $\boldsymbol{f}_{\delta}$ is typically very small
and thus changing the distribution does not make much difference.
In addition, after the coarse frequency synchronization, the residual
frequency offset is limited to a small range, suitable for the exponential
drop off of the Gaussian distribution. The pilot signals have average
power $\rho=\frac{1}{n}\text{Tr}\left(S^{\dagger}S\right)$. We consider
both the general case and the special case of orthogonal pilots where
$S^{\dagger}S=\frac{n\rho}{l_{\text{t}}}I_{l_{\text{t}}\times l_{\text{t}}}$.
{} %
{} %
{} %
{} %
{} %
{} %
{} 

\section{The Optimization Problem and Solution\label{sec:The-Optimization-Problem}}

To perform joint MAP estimation of channel and frequency offset, we
solve the following optimization problem. 
\begin{problem}
\label{prob:joint-fine-freq-channel} The problem of joint MAP estimation
of the fine frequency offset and the channel is
\begin{eqnarray}
 &  & (\hat{\vec{h}},\hat{f}_{\delta})\nonumber \\
 & = & \arg\max_{\vec{h},f_{\delta}}f_{\vec{\boldsymbol{h}},\boldsymbol{f}_{\delta},\vec{\boldsymbol{y}}}(\vec{h},f_{\delta},\vec{y})\nonumber \\
 & = & \arg\max_{\vec{h},f_{\delta}}f_{\vec{\boldsymbol{h}}|\vec{\boldsymbol{y}},\boldsymbol{f}_{\delta}}(\vec{h}|\vec{y},f_{\delta})f_{\boldsymbol{f}_{\delta},\vec{\boldsymbol{y}}}(f_{\delta},\vec{y})\nonumber \\
 & = & \arg\max_{f_{\delta}}\left(\arg\max_{\vec{h}}f_{\vec{\boldsymbol{h}}|\vec{\boldsymbol{y}},\boldsymbol{f}_{\delta}}(\vec{h}|\vec{y},f_{\delta})\right)\nonumber \\
 &  & \times f_{\vec{\boldsymbol{y}}|\boldsymbol{f}_{\delta}}(\vec{y}|f_{\delta})f_{\boldsymbol{f}_{\delta}}(f_{\delta}).\label{eq:MAP-h-f}
\end{eqnarray}
\end{problem}
\textbf{Solution:} The maximization over $f_{\delta}$ and $\vec{h}$
appears coupled but are actually separable, as shown in the following
steps.
\begin{enumerate}
\item Perform the MAP estimation of the channel given a frequency offset
$f_{\delta}$:
\begin{eqnarray}
\hat{\vec{h}}(\vec{y},f_{\delta}) & = & \arg\max_{\vec{h}}f_{\vec{\boldsymbol{h}}|\vec{\boldsymbol{y}},\boldsymbol{f}_{\delta}}(\vec{h}|\vec{y},f_{\delta}).\label{eq:max-h}
\end{eqnarray}
\item Substitute the above result in to estimate the CFO using
\begin{eqnarray*}
\hat{f}_{\delta} & = & \arg\max_{f_{\delta}}f_{\vec{\boldsymbol{h}}|\vec{\boldsymbol{y}},\boldsymbol{f}_{\delta}}(\hat{\vec{h}}(\vec{y},f_{\delta})|\vec{y},f_{\delta})\\
 &  & \times f_{\vec{\boldsymbol{y}}|\boldsymbol{f}_{\delta}}(\vec{y}|f_{\delta})f_{\boldsymbol{f}_{\delta}}(f_{\delta})\\
 & = & \arg\max_{f_{\delta}}f_{\vec{\boldsymbol{y}}|\boldsymbol{f}_{\delta}}(\vec{y}|f_{\delta})f_{\boldsymbol{f}_{\delta}}(f_{\delta}),
\end{eqnarray*}
 where we show below that $f_{\vec{\boldsymbol{h}}|\vec{\boldsymbol{y}},\boldsymbol{f}_{\delta}}(\hat{\vec{h}}(\vec{y},f_{\delta})|\vec{y},f_{\delta})$
is not a function of $f_{\delta}$. Therefore, the joint estimations
of frequency offset and channel are separable and we can solve an
individual MAP estimation of $\boldsymbol{f}_{\delta}$ with channel
state distribution information. If wanted, one can assume that $\boldsymbol{f}_{\delta}$
is uniform either over all real number or over a small interval, or
is Gaussian with infinite variance. Then, the MAP estimation of $\boldsymbol{f}_{\delta}$
can be converted to the ML estimation,
\begin{eqnarray*}
\hat{f}_{\delta} & = & \arg\max_{f_{\delta}}f_{\vec{\boldsymbol{y}}|\boldsymbol{f}_{\delta}}(\vec{y}|f_{\delta})f_{\boldsymbol{f}_{\delta}}(f_{\delta})\\
 & = & \arg\max_{f_{\delta}}f_{\vec{\boldsymbol{y}}|\boldsymbol{f}_{\delta}}(\vec{y}|f_{\delta}).
\end{eqnarray*}
\item Finally, $\hat{\vec{h}}(\vec{y},\hat{f}_{\delta})$ gives the solution
of the channel estimation.
\end{enumerate}

\subsection{MAP and ML Channel Estimation}

For the first step of the solution, the channel model implies $\vec{\boldsymbol{h}},\vec{\boldsymbol{y}}$
are jointly Gaussian conditioned on $\boldsymbol{f}_{\delta}$. Therefore,
\begin{eqnarray*}
f_{\vec{\boldsymbol{h}}|\vec{\boldsymbol{y}},\boldsymbol{f}_{\delta}}(\vec{h}|\vec{y},f_{\delta}) & = & \mathcal{CN}\left(\hat{\vec{h}}_{\text{MMSE}}(\vec{y},f_{\delta}),\Sigma_{\hat{\vec{\boldsymbol{h}}}_{\text{MMSE}}}\right)(\vec{h}),
\end{eqnarray*}
 where $\mathcal{CN}\left(\vec{\mu},\Sigma\right)(\vec{x})=\frac{1}{\det\left(\pi\Sigma\right)}e^{-\left(x-\vec{\mu}\right)^{\dagger}\Sigma^{-1}\left(x-\vec{\mu}\right)}$
denotes the circularly symmetric complex Gaussian density function;
$\hat{\vec{h}}_{\text{MMSE}}(\vec{y},f_{\delta})$ is the MMSE estimate
of $\vec{\boldsymbol{h}}$ and $\Sigma_{\hat{\vec{\boldsymbol{h}}}_{\text{MMSE}}}$
is the MMSE estimation error covariance, which does not depend on
$\vec{y}$ or $f_{\delta}$, as shown below.

To calculate $\hat{\vec{h}}_{\text{MMSE}}(\vec{y},f_{\delta})$, find
the mean of $\vec{\boldsymbol{y}}$ given the frequency offset as
\begin{eqnarray}
\vec{\mu}_{\vec{\boldsymbol{y}}|\boldsymbol{f}_{\delta}} & = & \text{E}\left[\vec{\boldsymbol{y}}|\{\boldsymbol{f}_{\delta}=f_{\delta}\}\right]\nonumber \\
 & = & \text{E}\left[\grave{X}\vec{\boldsymbol{h}}+\vec{\boldsymbol{n}}\right]\nonumber \\
 & = & \grave{X}\vec{\mu}_{\vec{\boldsymbol{h}}}.\label{eq:Ey-f}
\end{eqnarray}
By the MMSE estimation theory, the the MMSE estimate is %
\begin{eqnarray*}
\hat{\vec{h}}_{\text{MMSE}}(\vec{y},f_{\delta}) & = & \left(\grave{X}^{\dagger}\grave{X}+\Sigma_{\vec{\boldsymbol{h}}}^{-1}\right)^{-1}\grave{X}^{\dagger}\left(\vec{y}-\grave{X}\vec{\mu}_{\vec{\boldsymbol{h}}}\right)+\vec{\mu}_{\vec{\boldsymbol{h}}}\\
 & = & A\grave{X}^{\dagger}\left(\vec{y}-\grave{X}\vec{\mu}_{\vec{\boldsymbol{h}}}\right)+\vec{\mu}_{\vec{\boldsymbol{h}}}\\
 & = & A\grave{S}^{\dagger}\grave{F}^{\dagger}\vec{y}+\vec{b},
\end{eqnarray*}
where
\begin{eqnarray}
A & = & \left[\left[a_{r_{1},t_{1},r_{2},t_{2}}\right]_{t_{1},t_{2}}\right]_{r_{1},r_{2}}\nonumber \\
 & \triangleq & \begin{cases}
\left(\Sigma_{\vec{\boldsymbol{h}}}^{-1}+\frac{n\rho}{l_{\text{t}}}I\right)^{-1} & S^{\dagger}S=\frac{n\rho}{l_{\text{t}}}I_{l_{\text{t}}\times l_{\text{t}}}\\
\left(\grave{S}^{\dagger}\grave{S}+\Sigma_{\vec{\boldsymbol{h}}}^{-1}\right)^{-1} & \text{else}
\end{cases},\label{eq:def-A}
\end{eqnarray}
\begin{eqnarray}
\vec{b} & = & \left[\left[b_{r,t}\right]_{t}\right]_{r}\nonumber \\
 & \triangleq & \left(I-A\grave{X}^{\dagger}\grave{X}\right)\vec{\mu}_{\vec{\boldsymbol{h}}}\nonumber \\
 & = & \begin{cases}
\left(I-\frac{n\rho}{l_{\text{t}}}A\right)\vec{\mu}_{\vec{\boldsymbol{h}}} & S^{\dagger}S=\frac{n\rho}{l_{\text{t}}}I_{l_{\text{t}}\times l_{\text{t}}}\\
\left(I-AS^{\dagger}S\right)\vec{\mu}_{\vec{\boldsymbol{h}}} & \text{else}
\end{cases},\label{eq:def-b}
\end{eqnarray}
{} and the estimation error covariance matrix is %
{} 
\begin{eqnarray*}
\Sigma_{\hat{\vec{\boldsymbol{h}}}_{\text{MMSE}}} & = & A,
\end{eqnarray*}
 which is not a function of $f_{\delta}$. %
{} Consequently, the solution to (\ref{eq:max-h}) is
\begin{eqnarray*}
\hat{\vec{h}}(\vec{y},f_{\delta}) & = & \hat{\vec{h}}_{\text{MMSE}}(\vec{y},f_{\delta}).
\end{eqnarray*}
Then 
\begin{eqnarray*}
f_{\vec{\boldsymbol{h}}|\vec{\boldsymbol{y}},\boldsymbol{f}_{\delta}}(\hat{\vec{h}}(\vec{y},f_{\delta})|\vec{y},f_{\delta}) & = & \frac{1}{\det\left(\pi\Sigma_{\hat{\vec{\boldsymbol{h}}}_{\text{MMSE}}}\right)}
\end{eqnarray*}
 is not a function of $f_{\delta}$. 
\begin{rem}
Setting $\Sigma_{\vec{\boldsymbol{h}}}^{-1}=\boldsymbol{0}$ in the
above provides ML or least square channel estimation.
\end{rem}

\subsection{MAP and ML Frequency Offset Estimation}

For the second step, we observe that conditioned on $\{\boldsymbol{f}_{\delta}=f_{\delta}\}$,
$\vec{\boldsymbol{y}}$ is a summation of Gaussian random variables
and has distribution $\mathcal{CN}\left(\vec{\mu}_{\vec{\boldsymbol{y}}|\boldsymbol{f}_{\delta}}(f_{\delta}),\Sigma_{\vec{\boldsymbol{y}}|\boldsymbol{f}_{\delta}}(f_{\delta})\right)$,
where 
\begin{eqnarray}
\Sigma_{\vec{\boldsymbol{y}}|\boldsymbol{f}_{\delta}} & = & \grave{X}\Sigma_{\vec{\boldsymbol{h}}}\grave{X}^{\dagger}+I,\label{eq:cov-y}
\end{eqnarray}
 according to \prettyref{eq:channel-model-fine-freq}.%
{} Using identity $\det(I+AB)=\det(I+BA)$, we obtain 
\begin{eqnarray}
 &  & \det(\pi\Sigma_{\vec{\boldsymbol{y}}|\boldsymbol{f}_{\delta}})\nonumber \\
 & = & \left(\pi\right)^{nl_{\text{r}}}\det\left(I+\Sigma_{\vec{\boldsymbol{h}}}\grave{X}^{\dagger}\grave{X}\right)\nonumber \\
 & = & \begin{cases}
\left(\pi\right)^{nl_{\text{r}}}\det\left(I+\frac{\rho n}{l_{\text{t}}}\Sigma_{\vec{\boldsymbol{h}}}\right) & S^{\dagger}S=\frac{n\rho}{l_{\text{t}}}I_{l_{\text{t}}\times l_{\text{t}}}\\
\left(\pi\right)^{nl_{\text{r}}}\det\left(I+\Sigma_{\vec{\boldsymbol{h}}}\grave{S}^{\dagger}\grave{S}\right) & \text{else}
\end{cases},\label{eq:det-not-depend-on-f}
\end{eqnarray}
which is not a function of $f_{\delta}$. We have the following theorem.
\begin{thm}
\label{thm:MAP-f}For Gaussian distributed random channel $\vec{\boldsymbol{h}}$,
the MAP frequency offset estimate is 
\begin{eqnarray}
\hat{f}_{\delta} & = & \arg\max_{f_{\delta}}f_{\vec{\boldsymbol{y}}|\boldsymbol{f}_{\delta}}(\vec{y}|f_{\delta})f_{\boldsymbol{f}_{\delta}}(f_{\delta})\nonumber \\
 & = & \arg\max_{f_{\delta}}\frac{1}{\det(\pi\Sigma_{\vec{\boldsymbol{y}}|\boldsymbol{f}_{\delta}})}e^{-\left(\vec{y}-\vec{\mu}_{\vec{\boldsymbol{y}}|\boldsymbol{f}_{\delta}}\right)^{\dagger}\Sigma_{\vec{\boldsymbol{y}}|\boldsymbol{f}_{\delta}}^{-1}\left(\vec{y}-\vec{\mu}_{\vec{\boldsymbol{y}}|\boldsymbol{f}_{\delta}}\right)}\nonumber \\
 &  & \times\frac{1}{\sqrt{2\pi\sigma_{\boldsymbol{f}_{\delta}}^{2}}}e^{-\frac{1}{2}\left(f_{\delta}-\mu_{\boldsymbol{f}_{\delta}}\right)^{\dagger}\sigma_{\boldsymbol{f}_{\delta}}^{-2}\left(f_{\delta}-\mu_{\boldsymbol{f}_{\delta}}\right)}\label{eq:jointpdff}\\
 & = & \arg\max_{f_{\delta}}g(\vec{y},f_{\delta}),\label{eq:max-g}
\end{eqnarray}
 where 
\begin{eqnarray}
 &  & g(\vec{y},f_{\delta})\nonumber \\
 & \triangleq & 2\Re\left[\left\langle \grave{X}^{\dagger}\vec{y},\vec{b}\right\rangle \right]+\left(\grave{X}^{\dagger}\vec{y}\right)^{\dagger}A\left(\grave{X}^{\dagger}\vec{y}\right)\nonumber \\
 &  & -\frac{1}{2}\sigma_{\boldsymbol{f}_{\delta}}^{-2}\left|f_{\delta}-\mu_{\boldsymbol{f}_{\delta}}\right|^{2};\label{eq:gyf}
\end{eqnarray}
{} $A$ is given in \prettyref{eq:def-A} and $\vec{b}$ is given in
\prettyref{eq:def-b}, which are not functions of $f_{\delta}$; $\grave{X}$
is a function of $f_{\delta}$. The ML estimator is obtained by setting
$\sigma_{\boldsymbol{f}_{\delta}}^{-2}=0$ in \prettyref{eq:gyf}.
\end{thm}
The proof is given in Appendix \ref{sec:Proof-of-MAP}. When $f_{\boldsymbol{f}_{\delta}}(f_{\delta})$
is a uniform distribution, the MAP estimator becomes the ML estimator.
The uniform distribution is achieved by $\sigma_{\boldsymbol{f}_{\delta}}^{2}\rightarrow\infty$
and thus $\sigma_{\boldsymbol{f}_{\delta}}^{-2}\rightarrow0$. 

The above proves the following theorem on the separable solution.
\begin{thm}
The joint fine frequency offset and channel estimation Problem \ref{prob:joint-fine-freq-channel}
can be decomposed into two separable optimization problems:
\begin{enumerate}
\item The MAP estimation of $\boldsymbol{f}_{\delta}$ is 
\begin{eqnarray*}
\hat{f}_{\delta} & = & \arg\max_{f_{\delta}}f_{\vec{\boldsymbol{y}}|\boldsymbol{f}_{\delta}}(\vec{y}|f_{\delta})f_{\boldsymbol{f}_{\delta}}(f_{\delta})=\arg\max_{f_{\delta}}g(\vec{y},f_{\delta}).
\end{eqnarray*}
Setting $f_{\boldsymbol{f}_{\delta}}(f_{\delta})$ as a constant,
or making $\sigma_{\boldsymbol{f}_{\delta}}^{-2}=0$, it reduces to
the ML estimation of $\boldsymbol{f}_{\delta}$. %
\item MAP or MMSE estimation of $\vec{\boldsymbol{h}}$ given the above
$\hat{f}_{\delta}$ is%
\begin{eqnarray*}
\hat{\vec{h}}(\vec{y},\hat{f}_{\delta}) & = & \arg\max_{\vec{h}}f_{\vec{\boldsymbol{h}}|\vec{\boldsymbol{y}},\boldsymbol{f}_{\delta}}(\vec{h}|\vec{y},\hat{f}_{\delta})=\hat{\vec{h}}_{\text{MMSE}}(\vec{y},\hat{f}_{\delta}).
\end{eqnarray*}
\end{enumerate}
\end{thm}
\begin{rem}
Setting $\Sigma_{\vec{\boldsymbol{h}}}^{-1}=\boldsymbol{0}$ in the
above provides frequency estimation without prior knowledge on channel
as in ML estimation.
\end{rem}
The MMSE estimation of the channel is straightforward. We focus on
the frequency offset estimation algorithms.

\section{Fine Frequency Offset Estimation Algorithms\label{sec:Fine-Frequency-Synchronization}}

We design low complexity algorithms for frequency offset estimation
for the general case and for two special cases with different pilot
signal structures.

\subsection{General Case}

The intuitive meaning of the frequency offset estimation (\ref{eq:max-g})
is to find $f_{\delta}$ to de-rotate $\vec{y}$ so that its energy
projected to the signal space is maximized. We may do so by solving
$\frac{\partial g(\vec{y},f_{\delta})}{\partial f_{\delta}}=0$. It
is summarized in the following theorem.
\begin{thm}
\label{thm:dgdf-0}The optimal solution $f_{\delta}$ to the MAP estimation
problem satisfies
\begin{eqnarray}
0 & = & \frac{\partial g(\vec{y},f_{\delta})}{\partial f_{\delta}}\nonumber \\
 & = & -4\pi\Im\left[\sum_{k=1}^{n-1}e^{j2\pi f_{\delta}k}kr_{k}e^{-j\theta_{k}}\right]\nonumber \\
 &  & -\sigma_{\boldsymbol{f}_{\delta}}^{-2}\left(f_{\delta}-\mu_{\boldsymbol{f}_{\delta}}\right)\nonumber \\
 & = & -4\pi\sum_{k=1}^{n-1}kr_{k}\sin\left(2\pi k\left(f_{\delta}-\frac{\theta_{k}}{2\pi k}\right)\right)\nonumber \\
 &  & -\sigma_{\boldsymbol{f}_{\delta}}^{-2}\left(f_{\delta}-\mu_{\boldsymbol{f}_{\delta}}\right),\label{eq:dg-df-general}
\end{eqnarray}
where $r_{k}>0$ and
\begin{eqnarray}
r_{k}e^{-j\theta_{k}} & \triangleq & \sum_{r,t}s_{t,k+1}y_{r,k+1}^{*}b_{r,t}+\nonumber \\
 &  & \sum_{k_{1}=k+1}^{n}\sum_{r_{1},t_{1},r_{2},t_{2}}a_{r_{1},t_{1},r_{2},t_{2}}\times\nonumber \\
 &  & s_{t_{1},k_{1}}s_{t_{2},k_{1}-k}^{*}y_{r_{2},k_{1}-k}y_{r_{1},k_{1}}^{*}.\label{eq:rtheta-general}
\end{eqnarray}
\end{thm}
It is proved in Appendix \ref{sec:Cal-dgdf}.

To solve the nonlinear equation \prettyref{eq:dg-df-general}, we
observe the following. For high SNR, $\frac{\theta_{k}+m_{k}2\pi}{2\pi k}$
approaches $f_{\delta}$, where $m_{k}\in\mathbb{Z}$ is for phase
unwrapping. Therefore, the asymptotic optimal solution is to employ
$\sin(x)\doteq x$ to solve (\ref{eq:dg-df-general}) and obtain asymptotic
MAP estimate
\begin{eqnarray}
\hat{f}_{\delta} & \doteq & \frac{4\pi\sum_{k=1}^{n-1}kr_{k}(\theta_{k}+m_{k}2\pi)+\sigma_{\boldsymbol{f}_{\delta}}^{-2}\mu_{\boldsymbol{f}_{\delta}}}{8\pi^{2}\sum_{k=1}^{n-1}k^{2}r_{k}+\sigma_{\boldsymbol{f}_{\delta}}^{-2}},\label{eq:fhat-general}
\end{eqnarray}
where $\doteq$ is an asymptotic equality when the $\text{SNR}\rightarrow\infty$.
The solution is a weighted average of of $\theta_{k}$ and mean $\mu_{\boldsymbol{f}_{\delta}}$.

The above is summarized in Algorithm \ref{alg:estimate-freq-general}.

\begin{algorithm}[h]
\caption{\label{alg:estimate-freq-general}General Frequency Offset Estimation}

\begin{enumerate}
\item \textbf{Input: }Matched filter output $y_{r,k}$, $r=1,...,l_{\text{r}}$,
$k=1,...,n$.
\item Calculate $r_{k}e^{-j\theta_{k}}$,$k=1,...,n-1$, according to \prettyref{eq:rtheta-general}
\item $\left[\theta_{k}+m_{k}2\pi\right]_{k=1:n-1}$ $=\text{phase unwrap}\left(\left[\theta_{k}\right]_{k=1:n-1}\right)$
\item Calculate $\hat{f}_{\delta}$ according to \prettyref{eq:fhat-general}
\item \textbf{Output:} $\hat{f}_{\delta}$.
\end{enumerate}
\end{algorithm}

\begin{rem}
The ML estimation algorithm can be obtained by setting $\sigma_{\boldsymbol{f}_{\delta}}^{-2}=0$
in \prettyref{eq:fhat-general}.
\end{rem}
\begin{rem}
The algorithm is almost in closed form except for a phase unwrapping.
Thus, the complexity is very low.
\end{rem}
\begin{rem}
An alternative way to use $\mu_{\boldsymbol{f}_{\delta}}$ is to de-rotate
the received continuous time signals by $e^{-j2\pi\mu_{\boldsymbol{f}_{\delta}}(k-1)}y_{r,k}$
and then estimate the frequency offset by setting $\mu_{\boldsymbol{f}_{\delta}}=0$
in \prettyref{eq:fhat-general}. The advantage is to increase the
estimation range limit from $|\boldsymbol{f}_{\delta}|<0.5$ to $|\boldsymbol{f}_{\delta}-\mu_{\boldsymbol{f}_{\delta}}|<0.5$. 
\end{rem}
\begin{rem}
If we want to use a closed loop approach like phase lock loop, based
on (\ref{eq:dg-df-general}), we can use 
\begin{eqnarray*}
e & = & -\gamma\sum_{k=1}^{n-1}kr_{k}\sin\left(2\pi k\left(f_{\delta}-\frac{\theta_{k}}{2\pi k}\right)\right)
\end{eqnarray*}
 as the feedback error, where $\gamma$ is an appropriate step size.
This is equivalent to the smoothing filter approach when the filter
has feedback loops. 
\end{rem}
\begin{rem}
Our MAP estimation of channel and frequency offset can be employed
to deal with time varying cases. For example, if $\boldsymbol{f}_{\delta}$
is time varying from packet to packet, we can use current estimate,
the estimation error variance, to be calculated from the Bayesian
Cramer-Rao lower bound in Section \prettyref{sec:Performance-Analysis},
and the correlation between the current and the next frequency offset
to calculate the prior distribution of the next frequency offset.
The prior distribution then is used in the MAP estimation of the next
packet/frame's frequency offset.
\end{rem}
To obtain further insight of the effect of pilot signal structure
on frequency offset estimation, we consider zero mean i.i.d. channel
and two typical kinds of pilot signals, periodic pilot and time division
pilot, in the next two subsections. When channel covariance $\Sigma_{\vec{\boldsymbol{h}}}=\sigma_{\boldsymbol{h}}^{2}I$,
we have 
\begin{eqnarray*}
A & = & \left(\sigma_{\boldsymbol{h}}^{-2}+\frac{n\rho}{l_{\text{t}}}\right)^{-1}I,
\end{eqnarray*}
 and thus 
\begin{eqnarray}
a_{r_{1},t_{1},r_{2},t_{2}} & = & \left(\sigma_{\boldsymbol{h}}^{-2}+\frac{n\rho}{l_{\text{t}}}\right)^{-1}\delta[r_{1}-r_{2}]\delta[t_{1}-t_{2}].\label{eq:a-white-channel}
\end{eqnarray}
If $\vec{\mu}_{\vec{\boldsymbol{h}}}=\vec{0}$, then $\vec{b}=\vec{0}$
and $b_{r,t}=0$. Therefore, \ref{eq:rtheta-general} can be simplified
to 
\begin{eqnarray}
 &  & r_{k}e^{-j\theta_{k}}\nonumber \\
 & = & \left(\sigma_{\boldsymbol{h}}^{-2}+\frac{n\rho}{l_{\text{t}}}\right)^{-1}\sum_{k_{1}=k+1}^{n}\underset{\ddot{s}_{k_{1},k_{1}-k}}{\underbrace{\sum_{t}s_{t,k_{1}}s_{t,k_{1}-k}^{*}}}\nonumber \\
 &  & \sum_{r}y_{r,k_{1}-k}y_{r,k_{1}}^{*},\label{eq:rtheta-zero-channel}
\end{eqnarray}
 where 
\begin{eqnarray}
\ddot{s}_{k_{1},k_{2}} & \triangleq & \sum_{t}s_{t,k_{1}}s_{t,k_{2}}^{*}.\label{eq:def-sddot}
\end{eqnarray}

\subsection{Special Case: Scrambled Periodic Pilot and Zero Mean i.i.d. Channel}

We define \emph{Scrambled Periodic Pilot} as 
\begin{eqnarray}
S & = & \sqrt{\rho}\underset{C}{\underbrace{\left[c_{k}\right]_{k,k=1:n}}}\underset{\left[O\right]_{i=1:m}}{\underbrace{\left[I_{l_{\text{t}}}\right]_{i=1:m}O}},\label{eq:scrambled-pilot-2}
\end{eqnarray}
where $I_{l_{\text{t}}}$ is an $l_{\text{t}}\times l_{\text{t}}$
identity matrix; $n=ml_{\text{t}}$ is assumed for $m\in\mathbb{Z}^{+}$.
It has a structure of scrambled periodic matrix $\left[O\right]_{i=1:m}=\left[\begin{array}{c}
O\\
O\\
\vdots
\end{array}\right]\in\mathbb{C}^{n\times l_{\text{t}}}$, which is a block matrix with $m$ copies of an unitary matrix $O\in\mathbb{C}^{l_{\text{t}}\times l_{\text{t}}}$
on top of each other. Matrix $O$ satisfies $O^{\dagger}O=OO^{\dagger}=I_{l_{\text{t}}}$.
The scrambling code is $\vec{c}=\left[c_{k}\right]_{k=1:n}\in\mathbb{C}^{n\times1}$,
where $|c_{k}|=1,\ \forall k$. Diagonal matrix $C$'s diagonal elements
are from $\vec{c}$. A simple example for $c_{k}=1$, $O=I_{l_{\text{t}}}$,
$m=2$, $l_{\text{t}}=3$ is
\begin{eqnarray*}
S & = & \sqrt{\rho}\left[\begin{array}{ccc}
1 & 0 & 0\\
0 & 1 & 0\\
0 & 0 & 1\\
1 & 0 & 0\\
0 & 1 & 0\\
0 & 0 & 1
\end{array}\right].
\end{eqnarray*}
 Another example of this pilot structure is rows or columns of the
Hadamard matrix. The freedom of choosing $C$ and $O$ offers flexibility
for this structure. For example, $O$ could be a Hadamard matrix or
a Fourier transform matrix $\left[e^{-j2\pi\frac{ik}{l_{\text{t}}}}\right]_{i=1:l_{\text{t}},k=1:l_{\text{t}}}$,
while $C$ could be a Gold or a Zadoff-Chu sequence \cite{Chu_1972ITIT_PolyphaseCodesGoodPeriodicCorrelationPropertiesCorresp}.

Observe that for periodic pilot, in \prettyref{eq:rtheta-zero-channel},
\begin{eqnarray*}
\ddot{s}_{k_{1},k_{1}-k} & = & \rho c_{k_{1}}c_{k_{1}-k}^{*}
\end{eqnarray*}
 is only nonzero for $k=il_{\text{t}}$, $i=1,...,m-1$. Define $k_{1}=(i_{1}-1)l_{\text{t}}+i_{2}$,
$i_{1}=i+1,...,m$, $i_{2}=1,...,l_{\text{t}}$, to simplify \prettyref{eq:rtheta-zero-channel}
to 
\begin{eqnarray}
 &  & r_{il_{\text{t}}}e^{-j\theta_{il_{\text{t}}}}\nonumber \\
 & = & \left(\sigma_{\boldsymbol{h}}^{-2}+\frac{n\rho}{l_{\text{t}}}\right)^{-1}\rho\sum_{i_{1}=i+1}^{m}\sum_{r}\sum_{i_{2}=1}^{l_{\text{t}}}c_{(i_{1}-1-i)l_{\text{t}}+i_{2}}^{*}\nonumber \\
 &  & \times y_{r,(i_{1}-1-i)l_{\text{t}}+i_{2}}c_{(i_{1}-1)l_{\text{t}}+i_{2}}y_{r,(i_{1}-1)l_{\text{t}}+i_{2}}^{*}.\label{eq:rtheta-periodic}
\end{eqnarray}
Consequently, \prettyref{eq:fhat-general} is simplified to
\begin{eqnarray}
\hat{f}_{\delta} & \doteq & \frac{4\pi\sum_{i=1}^{m-1}il_{\text{t}}r_{il_{\text{t}}}(\theta_{il_{\text{t}}}+m_{il_{\text{t}}}2\pi)+\sigma_{\boldsymbol{f}_{\delta}}^{-2}\mu_{\boldsymbol{f}_{\delta}}}{8\pi^{2}\sum_{i=1}^{m-1}(il_{\text{t}})^{2}r_{il_{\text{t}}}+\sigma_{\boldsymbol{f}_{\delta}}^{-2}}.\label{eq:fhat-periodic}
\end{eqnarray}

Thus, the frequency estimation algorithm can be modified to Algorithm
\ref{alg:Fine-Frequency-Estimation-Periodic-2}.

\begin{algorithm}[h]
\caption{\label{alg:Fine-Frequency-Estimation-Periodic-2}Frequency Offset
Estimation for Scrambled Periodic Pilot and Zero Mean i.i.d. Channel}

\begin{enumerate}
\item \textbf{Input: }Matched filter output $y_{r,k}$, $r=1,...,l_{\text{r}}$,
$k=1,...,n$.
\item Calculate $r_{il_{\text{t}}}e^{-j\theta_{il_{\text{t}}}}$,$i=1,...,m-1$,
according to \prettyref{eq:rtheta-periodic}
\item $\left[\theta_{il_{\text{t}}}+m_{il_{\text{t}}}2\pi\right]_{i=1:m-1}$
$=\text{phase unwrap}\left(\left[\theta_{il_{\text{t}}}\right]_{i=1:m-1}\right)$
\item Calculate $\hat{f}_{\delta}$ according to \prettyref{eq:fhat-periodic}
\item \textbf{Output:} $\hat{f}_{\delta}$.
\end{enumerate}
\end{algorithm}

\subsection{Special Case: Scrambled Time Division Pilot and Zero Mean i.i.d.
Channel}

Another typical pilot signal used in practice is the \emph{Time Division
(TD)Pilot} 
\begin{eqnarray}
S & = & \sqrt{\rho}\underset{C}{\underbrace{\left[c_{k}\right]_{k,k=1:n}}}\left[\vec{1}_{m}\right]_{i,i=1:l_{\text{t}}},\label{eq:TDD-pilot-1}
\end{eqnarray}
where $n=ml_{\text{t}}$; only the first transmit antenna transmits
scrambled $m$ ones, followed by that only the second antenna transmits
$m$ scrambled ones, \emph{etc.}. Vector $\vec{1}_{m}$ has $m$ ones
on top of each other. Diagonal block matrix $\left[\vec{1}_{m}\right]_{i,i=1:l_{\text{t}}}=\left[\begin{array}{ccc}
\vec{1}_{m} & \vec{0} & \ldots\\
\vec{0} & \vec{1}_{m} & \cdots\\
\vdots & \vdots & \ddots
\end{array}\right]\in\mathbb{R}^{n\times l_{\text{t}}}$. A simple example for $c_{k}=1$, $m=2$, $l_{\text{t}}=3$ is
\begin{eqnarray*}
S & = & \sqrt{\rho}\left[\begin{array}{ccc}
1 & 0 & 0\\
1 & 0 & 0\\
0 & 1 & 0\\
0 & 1 & 0\\
0 & 0 & 1\\
0 & 0 & 1
\end{array}\right].
\end{eqnarray*}

Observe that for time division pilot, in \prettyref{eq:rtheta-zero-channel},
\begin{eqnarray*}
\ddot{s}_{k_{1},k_{1}-k} & = & \rho c_{k_{1}}c_{k_{1}-k}^{*}
\end{eqnarray*}
 is nonzero for $k=i=1,...,m-1$. Define $k_{1}=(i_{2}-1)m+i_{1}$,
$i_{2}=1,...,l_{\text{t}}$, $i_{1}=i+1,...,m$, to simplify \prettyref{eq:rtheta-zero-channel}
to 
\begin{eqnarray}
 &  & r_{i}e^{-j\theta_{i}}\nonumber \\
 & = & \left(\sigma_{\boldsymbol{h}}^{-2}+\frac{n\rho}{l_{\text{t}}}\right)^{-1}\rho\sum_{i_{1}=i+1}^{m}\sum_{r}\sum_{i_{2}=1}^{l_{\text{t}}}c_{(i_{2}-1)m+i_{1}-i}^{*}\nonumber \\
 &  & \times y_{r,(i_{2}-1)m+i_{1}-i}c_{(i_{2}-1)m+i_{1}}y_{r,(i_{2}-1)m+i_{1}}^{*}.\label{eq:rtheta-TD}
\end{eqnarray}
Consequently, \prettyref{eq:fhat-general} is simplified to
\begin{eqnarray}
\hat{f}_{\delta} & \doteq & \frac{4\pi\sum_{i=1}^{m-1}ir_{i}(\theta_{i}+m_{i}2\pi)+\sigma_{\boldsymbol{f}_{\delta}}^{-2}\mu_{\boldsymbol{f}_{\delta}}}{8\pi^{2}\sum_{i=1}^{m-1}i^{2}r_{i}+\sigma_{\boldsymbol{f}_{\delta}}^{-2}}.\label{eq:fhat-TD}
\end{eqnarray}

Thus, the frequency estimation algorithm can be modified to Algorithm
\ref{alg:Fine-Frequency-Estimation-TDD-1}.

\begin{algorithm}[h]
\caption{\label{alg:Fine-Frequency-Estimation-TDD-1}Frequency Offset Estimation
for Scrambled Time Division Pilot and Zero Mean i.i.d. Channel}

\begin{enumerate}
\item \textbf{Input: }Matched filter output $y_{r,k}$, $r=1,...,l_{\text{r}}$,
$k=1,...,n$.
\item Calculate $r_{i}e^{-j\theta_{i}}$,$i=1,...,m-1$, according to \prettyref{eq:rtheta-TD}
\item $\left[\theta_{i}+m_{i}2\pi\right]_{i=1:m-1}$ $=\text{phase unwrap}\left(\left[\theta_{i}\right]_{i=1:m-1}\right)$
\item Calculate $\hat{f}_{\delta}$ according to \prettyref{eq:fhat-TD}
\item \textbf{Output:} $\hat{f}_{\delta}$.
\end{enumerate}
\end{algorithm}

---------------

We observe that the both $\left[I_{l_{\text{t}}}\right]_{i=1:m}$
and $\left[\vec{1}_{m}\right]_{i,i=1:l_{\text{t}}}$ have one $1$
per row and $m$ 1's per column. They represent two opposite ways
to arrange the rows and are useful in different scenarios and have
different performance. The periodic structure is useful when we do
not want to switch on and off antennas. For the same amount of signal
energy, it only requires $\frac{1}{l_{\text{t}}}$ peak power per
antenna of the time division structure, because all antennas are on
all the time. The TD structure is useful when we need backward compatibility
to single antenna systems and when we can afford larger peak power
per antenna.

{} %

\section{Performance Bounds\label{sec:Performance-Analysis}}

We compare the mean square error of the above MAP frequency estimation
with Bayesian Cramér-Rao Lower Bound (BCRLB). We see below that when
$\sigma_{\boldsymbol{f}_{\delta}}^{-2}=0$, the BCRLB becomes CRLB
for mean square error conditioned on $\{\boldsymbol{f}_{\delta}=f_{\delta}\}$.
The bounds are not a function of $f_{\delta}$. Since the optimal
pilot for channel estimation is orthogonal across transmit antennas,
we assume such in the following.

The BCRLB is given in \cite{Tian_2013_DetectionEstimationModulationTheoryPartIDetectionEstimationFilteringTheory}
for parameter estimation with prior knowledge. The proof of the following
theorem implies that $\frac{\partial\ln\left(f_{\vec{\boldsymbol{y}}|\boldsymbol{f}_{\delta}}(\vec{y}|f_{\delta})f_{\boldsymbol{f}_{\delta}}(f_{\delta})\right)}{\partial f_{\delta}}$
and $\frac{\partial^{2}\ln\left(f_{\vec{\boldsymbol{y}}|\boldsymbol{f}_{\delta}}(\vec{y}|f_{\delta})f_{\boldsymbol{f}_{\delta}}(f_{\delta})\right)}{\partial f_{\delta}^{2}}$
are absolutely integrable with respect to $\vec{y}$ and $f_{\delta}$,
satisfying the conditions of BCRLB. 
\begin{thm}
\label{thm:CRLB} For any estimator satisfying 
\begin{eqnarray*}
\lim_{f_{\delta}\rightarrow\infty}\text{E}\left[\hat{\boldsymbol{f}}_{\delta}-f_{\delta}|\{\boldsymbol{f}_{\delta}=f_{\delta}\}\right]f_{\boldsymbol{f}_{\delta}}(f_{\delta}) & = & 0
\end{eqnarray*}
 and
\begin{eqnarray*}
\lim_{f_{\delta}\rightarrow-\infty}\text{E}\left[\hat{\boldsymbol{f}}_{\delta}-f_{\delta}|\{\boldsymbol{f}_{\delta}=f_{\delta}\}\right]f_{\boldsymbol{f}_{\delta}}(f_{\delta}) & = & 0,
\end{eqnarray*}
the mean square frequency estimation error for channel model (\ref{eq:channel-model-fine-freq})
with any orthogonal pilot signal $S$, satisfying $S^{\dagger}S=\frac{n\rho}{l_{\text{t}}}I$,
is lower bounded by the Bayesian CRLB:
\begin{eqnarray}
 &  & \text{E}\left[\left(\hat{\boldsymbol{f}}_{\delta}-\boldsymbol{f}_{\delta}\right)^{2}\right]\nonumber \\
 & \ge & \text{BCRLB}\nonumber \\
 & = & \frac{1}{-\text{E}\left[\frac{\partial^{2}\ln\left(f_{\vec{\boldsymbol{y}}|\boldsymbol{f}_{\delta}}(\vec{\boldsymbol{y}}|\boldsymbol{f}_{\delta})f_{\boldsymbol{f}_{\delta}}(\boldsymbol{f}_{\delta})\right)}{\partial\boldsymbol{f}_{\delta}^{2}}\right]},\label{eq:1-E}\\
 & = & \frac{1}{\beta+\sigma_{\boldsymbol{f}_{\delta}}^{-2}}.\label{eq:BCRLB}
\end{eqnarray}
Setting $\sigma_{\boldsymbol{f}_{\delta}}^{-2}=0$, the conditional
mean square error is lower bounded by the CRLB:
\begin{eqnarray}
 &  & \text{E}\left[\left(\hat{\boldsymbol{f}}_{\delta}-\boldsymbol{f}_{\delta}\right)^{2}|\{\boldsymbol{f}_{\delta}=f_{\delta}\}\right]\nonumber \\
 & \ge & \text{CRLB}\nonumber \\
 & = & \frac{1}{-\text{E}_{\vec{\boldsymbol{y}}|\{\boldsymbol{f}_{\delta}=f_{\delta}\}}\left[\frac{\partial^{2}\ln\left(f_{\vec{\boldsymbol{y}}|\boldsymbol{f}_{\delta}}(\vec{\boldsymbol{y}}|f_{\delta})\right)}{\partial f_{\delta}^{2}}\right]},\label{eq:1-E-1}\\
 & = & \frac{1}{\beta},\nonumber 
\end{eqnarray}
where 
\begin{eqnarray}
\beta & = & 8\pi^{2}\Re\left[\sum_{k=1}^{n-1}k^{2}\times\right.\nonumber \\
 &  & \left(\sum_{r,t}\sum_{t'}s_{t,k+1}s_{t',k+1}^{*}\mu_{\boldsymbol{h}_{r,t'}}^{*}b_{r,t}+\right.\nonumber \\
 &  & \sum_{k_{1}=k+1}^{n}\sum_{r_{1},t_{1},r_{2},t_{2}}a_{r_{1},t_{1},r_{2},t_{2}}\times\nonumber \\
 &  & s_{t_{1},k_{1}}s_{t_{2},k_{1}-k}^{*}\sum_{t_{2}'}s_{t_{2}',k_{1}-k}\sum_{t_{1}'}s_{t_{1}',k_{1}}^{*}\times\nonumber \\
 &  & \left.\left.\left(c_{\boldsymbol{h}_{r_{1},t_{1}'},\boldsymbol{h}_{r_{2},t_{2}'}}^{*}+\mu_{\boldsymbol{h}_{r_{2},t_{2}'}}\mu_{\boldsymbol{h}_{r_{1},t_{1}'}}^{*}\right)\right)\right]\label{eq:-E-general}
\end{eqnarray}
$a_{r_{1},t_{1},r_{2},t_{2}}$ is given in \prettyref{eq:def-A};
$b_{r,t}$ is given in \prettyref{eq:def-b}.
\begin{itemize}
\item If the channel is i.i.d. zero mean with $\Sigma_{\vec{\boldsymbol{h}}}=\sigma_{\boldsymbol{h}}^{2}I$
and $\vec{\mu}_{\vec{\boldsymbol{h}}}=\vec{0}$ and the pilot signals
are orthogonal, i.e., $S^{\dagger}S=\frac{n\rho}{l_{\text{t}}}I_{l_{\text{t}}\times l_{\text{t}}}$,
then 
\begin{eqnarray}
\beta & = & 8\pi^{2}l_{\text{r}}\left(\sigma_{\boldsymbol{h}}^{-2}+\frac{n\rho}{l_{\text{t}}}\right)^{-1}\sigma_{\boldsymbol{h}}^{2}\times\nonumber \\
 &  & \sum_{k=1}^{n-1}k^{2}\sum_{k_{1}=k+1}^{n}\left|\ddot{s}_{k_{1},k_{1}-k}\right|^{2},\label{eq:-E-zero-channel}
\end{eqnarray}
where $\ddot{s}_{k_{1},k_{1}-k}$ is defined in \prettyref{eq:def-sddot}.%

\begin{itemize}
\item In addition, for the periodic pilot signal in (\ref{eq:scrambled-pilot-2}),
{} 
\begin{eqnarray}
\beta=\beta_{\text{P}} & = & \frac{2}{3}\pi^{2}l_{\text{r}}l_{\text{t}}\left(1+\left(\frac{n\rho}{l_{\text{t}}}\sigma_{\boldsymbol{h}}^{2}\right)^{-1}\right)^{-1}\times\nonumber \\
 &  & \left(\frac{n\rho}{l_{\text{t}}}\sigma_{\boldsymbol{h}}^{2}\right)\left(n^{2}\left(1-\frac{l_{\text{t}}^{2}}{n^{2}}\right)\right)\label{eq:-E-zero-channel-periodic}
\end{eqnarray}
\item In addition, for the time division pilot signal in (\ref{eq:TDD-pilot-1}),
\begin{eqnarray}
\beta=\beta_{\text{T}} & = & l_{\text{t}}^{-2}\beta_{\text{P}}\label{eq:eq:-E-zero-channel-TD}
\end{eqnarray}
\end{itemize}
\end{itemize}
\end{thm}
The proof is given in Appendix \ref{sec:Proof-CRLB}.
\begin{rem}
\end{rem}
\begin{rem}
The CRLBs decrease with received signal SNR $\rho\sigma_{\boldsymbol{h}}^{2}$
in the order of $O\left(\frac{1}{\rho\sigma_{\boldsymbol{h}}^{2}}\right)$.
It decrease with the pilot length $n$ in the order of $O\left(\frac{1}{n^{3}}\right)$
and decrease with the number of receive antenna in the order of $O\left(\frac{1}{l_{\text{r}}}\right)$.
\end{rem}

\subsection{Pilot/Training Signal Design for CFO and Channel Estimation}

\emph{Orthogonality:} The BCRLB can guide the design of the pilot
signals for frequency estimation. The pilot signal is also used for
channel estimation. Since in general it is not practical to design
pilot signals for each specific channel correlation, one should design
it for i.i.d. channel coefficients. It is easy to prove that the optimal
pilot for channel estimation for i.i.d. channel satisfies $S^{\dagger}S=\frac{n\rho}{l_{\text{t}}}I$,
as long as $n\ge l_{\text{t}}$ so that the pilot signals are orthogonal
across transmit antennas. Therefore, the BCRLB with $\beta$ in \prettyref{eq:-E-zero-channel}
is the right one to guide the pilot signal design.

\emph{Time Spread:} To minimize the BCRLB, we need to maximize the
weighted sum $\sum_{k=1}^{n-1}k^{2}\sum_{k_{1}=k+1}^{n}\left|\ddot{s}_{k_{1},k_{1}-k}\right|^{2}$,
where $\ddot{s}_{k_{1},k_{1}-k}$ of \prettyref{eq:def-sddot} is
the inner product of the rows of $S$. The larger the row index difference,
the larger the weight $k^{2}$ is. This suggests to spread the energy
of the pilot signal at the top few and the bottom few rows of $S$,
with zeros in between and repeated rows at the top and the bottom.
For example, $S^{T}=\left[\begin{array}{ccccccc}
1 & 1 & 0 & \cdots & 0 & 1 & 1\\
1 & -1 & 0 & \cdots & 0 & -1 & 1
\end{array}\right]$ would be a good choice. The intuition is that the larger the spread,
the larger phase the frequency produces and thus, the easier to detect.
Another consideration is the acquisition range limited by the ambiguity
due to that $e^{j\theta}$ is a periodic function. Thus, the design
guide line of the pilot signals is to place it at the beginning and
repeat it at the end of a packet with enough consecutive symbol time
of pilot to satisfies the acquisition range requirement.

\emph{Periodic and Time Division Structures:} We observe that the
CRLB of periodic pilot, $\text{CRLB}_{\text{P}}=l_{\text{t}}^{-2}\text{CRLB}_{\text{T}}$,
has an $l_{\text{t}}^{2}$ advantage over the CRLB of the time division
pilot, $\text{CRLB}_{\text{T}}$, due to wider spreading of ones over
time in \prettyref{eq:scrambled-pilot-2}, resulting in $l_{\text{t}}$
times larger phase change for the same frequency offset. On the other
hand, the consecutive symbols in time division pilot results in larger
acquisition range. Combining both periodic and time division structures
in one pilot signal is expected to obtain the advantages of both,
as demonstrated in the next section.

\section{Simulation Results\label{sec:Simulation-Results}}

\emph{Summary of Observations:} We show the simulation results on
the CFO estimation. The MMSE channel estimation is standard and is
not shown. (1) We compare the average CFO MAP estimation square error
and BCRLB with the results of ML estimation and CRLB. Unlike the ML
estimation which diverges away from the CRLB at low SNR, the MAP estimation
achieves the BCRLB at almost all SNR range. (2) We consider three
kinds of pilot signals, periodic pilot, time division (TD) pilot,
and a combination of periodic and TD pilots. The periodic pilot is
shown to achieve the smaller BCRLB than the TD pilot, while the TD
pilot achieves larger acquisition range than the periodic pilot. The
combined pilot achieves the advantages of both periodic and TD pilots.
(3) When the CFO varies from packet to packet but is correlated, it
is shown that, unlike the ML estimation, the MAP estimation can track
the CFO and achieves much better performance.

\emph{Simulation Parameters:} (1) MIMO size: number of transmitter
antennas is $l_{t}=2$, number of receiver antennas is $l_{r}=2$.
(2) Pilot length: $n=16$ symbols; (3) CFO distribution: $\boldsymbol{f}_{\delta}$$\sim\mathcal{N}(\mu_{\boldsymbol{f}_{\delta}},\sigma_{\boldsymbol{f}_{\delta}}^{2})$
where $\mu_{\boldsymbol{f}_{\delta}}=0.01$, $\sigma_{\boldsymbol{f}_{\delta}}^{2}=10^{-5}$.

\subsection{Average Square Error and BCRLB\label{subsec:Mean-square-error}}

\begin{figure}
\begin{centering}
\subfloat[\label{fig:MSE_SNR_iid}i.i.d Gaussian channels]{\begin{centering}
\includegraphics[scale=0.5]{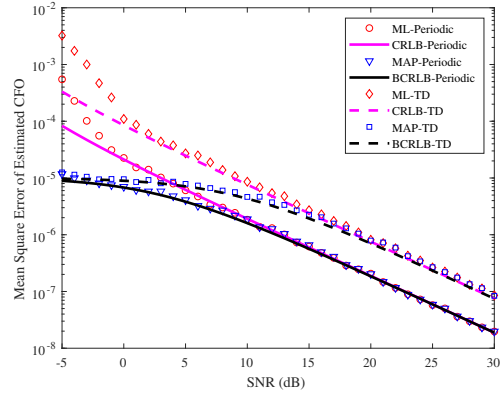}
\par\end{centering}
}
\par\end{centering}
\begin{centering}
\subfloat[\label{fig:MSE_SNR_Corr}Correlated non-zero mean channels]{\begin{centering}
\includegraphics[scale=0.5]{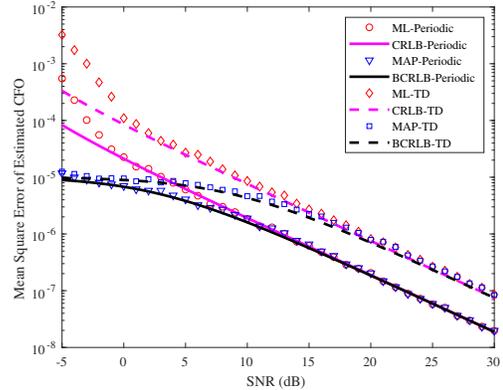}
\par\end{centering}
}
\par\end{centering}
\caption{\label{fig:MSE_SNR}Average estimation square error compared with
BCRLB/CRLB for ML and MAP estimation.}
\end{figure}

\emph{Approximately Achieving BCRLB:} Figure \ref{fig:MSE_SNR} shows
the average CFO MAP and ML estimation square errors and the BCRLB
and CRLB bounds for periodic and time division pilots and for i.i.d.
zero mean channels and spatially correlated non-zero mean channels.
It can be seen that at low SNR, the MAP results still almost overlap
with the BCRLB, which is not the case for the ML results. %
{} %
{} At high SNR, the average square error and the BCRLB/CRLB of the TD
pilot is $l_{t}^{2}=4$ times of that of periodic pilot.%

\subsection{Acquisition Range and Combined Pilot Structure\label{subsec:CFO-acquisition-range}}

\begin{figure}
\begin{centering}
\includegraphics[scale=0.5]{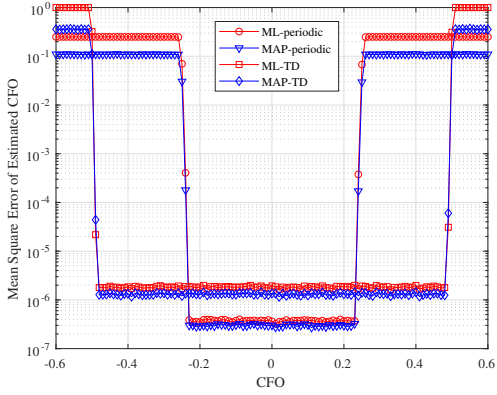}
\par\end{centering}
\caption{\label{fig:MSE-range}Acquisition range of CFO estimation for different
pilots.}
\end{figure}

\emph{Periodic and TD Pilots:} Figure \ref{fig:MSE-range} shows the
acquisition range of periodic and time division pilots at 20dB SNR.
One can observe that the periodic pilot has smaller square error while
the TD pilot has larger acquisition range that almost is the largest
possible of $\left|\boldsymbol{f}_{\delta}-\mu_{\boldsymbol{f}_{\delta}}\right|<0.5$.

\begin{figure}
\begin{centering}
\subfloat[\label{fig:MSE_SNR_Comb}Average square estimation errors and the
BCRLB bounds]{\begin{centering}
\includegraphics[scale=0.5]{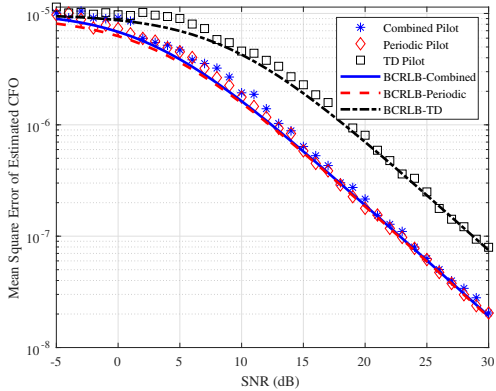}
\par\end{centering}
}
\par\end{centering}
\begin{centering}
\subfloat[\label{fig:Range_Comb}Acquisition ranges]{\begin{centering}
\includegraphics[scale=0.5]{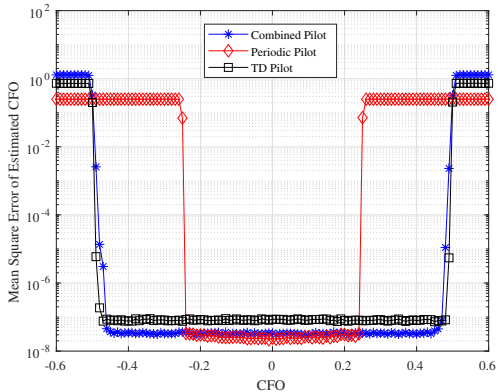}
\par\end{centering}
}
\par\end{centering}
\caption{\label{fig:Combined}MSE and acquisition range of combined pilot comparing
with the periodic and TD pilots}
\end{figure}

\emph{Combined Pilots:} This observation motivates the combination
of both pilot structure to design a pilot that has the advantages
of both. Figure \ref{fig:Combined} shows that this is indeed possible.
The combined pilot with 8 symbol time of periodic pilot followed by
8 symbol time of TD pilot achieves almost as small square error as
the periodic pilot of the same length%
{} and almost as large acquisition range as the TD pilot of the same
length. %

\subsection{MAP Estimation for CFO Tracking}

\emph{Tracking:} The MAP estimation provides a means for tracking
time varying parameters. Here, we give an example of CFO tracking
by taking advantage of the prior knowledge, where the CFO changes
from packet/frame to packet/frame but is correlated from frame to
frame. The estimated CFO $\hat{f}_{\delta,\nu}$ of the $\nu$-th
frame and its $\text{BCRLB}_{\nu}$ can be used together with the
correlation model to calculate the $\nu+1$-th frame's prior knowledge
of $\boldsymbol{f}_{\delta,\nu+1}$. In this example, we assume the
channel is independent from frame to frame to isolate the benefit
of CFO tracking.

\emph{Example Model:} We use a simple AR model for the CFO. It is
straightforward for a designer to adapt the result here for other
desired models. The model is $\boldsymbol{f}_{\delta,\nu+1}=\rho\boldsymbol{f}_{\delta,\nu}+(1-\rho)\mu_{\boldsymbol{f}_{\delta}}+\boldsymbol{w}_{\nu+1},$
where $0\le\rho\le1$ controls the correlation; $\mu_{\boldsymbol{f}_{\delta}}$
is the stationary mean; and $\boldsymbol{w}_{\nu+1}\sim\mathcal{N}(0,\sigma_{\boldsymbol{w}}^{2})$
is an i.i.d. Gaussian noise. Since the MAP estimation approximately
achieves BCRLB for almost all SNR according to the above simulation
results, we may approximately assume $B_{\nu}=\left\{ \boldsymbol{f}_{\delta,\nu}:\text{E}\left[\boldsymbol{f}_{\delta,\nu}\right]=\hat{f}_{\delta,\nu},\text{Var}\left[\boldsymbol{f}_{\delta,\nu}-\hat{f}_{\delta,\nu}\right]=\text{BCRLB}_{\nu}\right\} $
after finishing the estimation using frame $\nu$. Then according
to the AR model, the conditional mean $\text{E}\left[\boldsymbol{f}_{\delta,\nu+1}|B_{\nu}\right]$
$=\rho\hat{f}_{\delta,\nu}+(1-\rho)\mu_{\boldsymbol{f}_{\delta}}$
and conditional variance $\text{Var}\left[\boldsymbol{f}_{\delta,\nu+1}|B_{\nu}\right]$
$=\rho^{2}\text{BCRLB}_{\nu}+\sigma_{w}^{2}$ may serve as the prior
knowledge for frame $\nu+1$. The stationary variance is $\text{Var}\left[\boldsymbol{f}_{\delta,\infty}\right]$
$=\frac{\sigma_{w}^{2}}{1-\rho^{2}}$.

\emph{Observation:} For the AR model with $\rho=0.9,\mu_{\boldsymbol{f}_{\delta}}=0.1,\sigma_{w}^{2}=10^{-8},$%
{} $\text{SNR}=10\text{dB}$, and pilot length per frame $n=16$, the
simulation result is in Figure \ref{fig:MSE-of-AR1}. For the first
frame, the variance of the CFO is assumed to be infinity, resulting
in an ML estimation. The MAP estimation is applied since the 2nd frame.
We can observe that the MAP tracking performance improves over time
and is much better than the ML estimation that does not use the prior
knowledge. The BCRLBs for periodic and TD pilot in the figure overlap
in this case and assume perfect estimation of the previous frame.
Thus, it is a lower bound. If desired, the performance can be improved
by a backward belief propagation from the last frame to the first
frame.

\begin{center}
\begin{figure}
\begin{centering}
\includegraphics[scale=0.5]{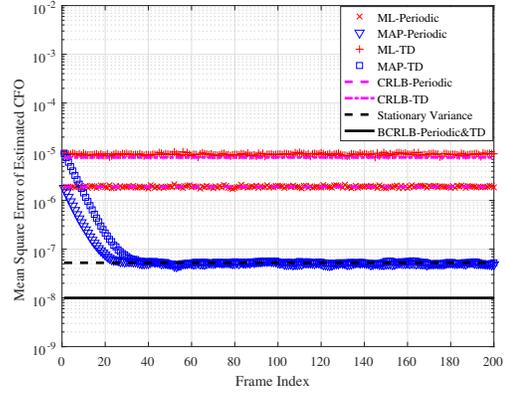}
\par\end{centering}
\caption{\label{fig:MSE-of-AR1}Tracking performance of CFO estimation in i.i.d
channels}
\end{figure}
\par\end{center}

\section{Conclusion\label{sec:Conclusion}}

In this paper, the solution of the joint MAP estimation of channel
states and the frequency offset is provided. Unexpectedly, the solution
is separable with an individual MAP estimation of the CFO with channel
statistic information first. An almost closed form algorithm is given.
The Bayesian Cramér-Rao Lower bounds (BCRLB) is derived in closed
form for the frequency offset estimation with prior knowledge. Based
on it, pilot signal signal design guideline is provided on mean square
error and acquisition range trade-off. Simulations with different
pilot structures are conducted and analyzed. The simulation results
show that the proposed algorithm has bound-approaching performance
at almost all SNR and a wide acquisition range. The MAP estimation
provides a different means to track time varying CFO, as demonstrated
in simulation, and can achieve much better performance than the ML
estimation.

\appendices{}

\section{\label{sec:Proof-of-MAP}Proof of Theorem \ref{thm:MAP-f} of the
MAP Estimator}

Since $\det(\pi\Sigma_{\vec{\boldsymbol{y}}|\boldsymbol{f}_{\delta}})$
is not a function of $f_{\delta}$, we can maximize the exponent in
\prettyref{eq:jointpdff} as
\begin{eqnarray}
\hat{f}_{\delta} & = & \arg\max_{f_{\delta}}f_{\vec{\boldsymbol{y}}|\boldsymbol{f}_{\delta}}(\vec{y}|f_{\delta})f_{\boldsymbol{f}_{\delta}}(f_{\delta})\nonumber \\
 & = & \arg\max_{f_{\delta}}-\left(\vec{y}-\vec{\mu}_{\vec{\boldsymbol{y}}|\boldsymbol{f}_{\delta}}\right)^{\dagger}\Sigma_{\vec{\boldsymbol{y}}|\boldsymbol{f}_{\delta}}^{-1}\left(\vec{y}-\vec{\mu}_{\vec{\boldsymbol{y}}|\boldsymbol{f}_{\delta}}\right)\nonumber \\
 &  & -\frac{1}{2}\sigma_{\boldsymbol{f}_{\delta}}^{-2}\left|f_{\delta}-\mu_{\boldsymbol{f}_{\delta}}\right|^{2}\label{eq:pdf-exponent}
\end{eqnarray}
{} 

We calculate the terms in \prettyref{eq:pdf-exponent} below. %
The conditional covariance 
\begin{eqnarray}
\Sigma_{\vec{\boldsymbol{y}}|\boldsymbol{f}_{\delta}}^{-1} & = & \left(\grave{X}\Sigma_{\vec{\boldsymbol{h}}}\grave{X}^{\dagger}+I\right)^{-1}\nonumber \\
 & = & \left(I-\grave{X}\left(\Sigma_{\vec{\boldsymbol{h}}}^{-1}+\grave{X}^{\dagger}\grave{X}\right)^{-1}\grave{X}^{\dagger}\right)\label{eq:by-woodbury-1}\\
 & = & \left(I-\grave{X}\left(\Sigma_{\vec{\boldsymbol{h}}}^{-1}+\grave{S}^{\dagger}\grave{S}\right)^{-1}\grave{X}^{\dagger}\right)\nonumber 
\end{eqnarray}
is converted by the Woodbury matrix identity. Then
\begin{eqnarray}
 &  & \left(\vec{y}-\vec{\mu}_{\vec{\boldsymbol{y}}|\boldsymbol{f}_{\delta}}\right)^{\dagger}I\left(\vec{y}-\vec{\mu}_{\vec{\boldsymbol{y}}|\boldsymbol{f}_{\delta}}\right)\nonumber \\
 & = & \vec{y}^{\dagger}\vec{y}+\vec{\mu}_{\vec{\boldsymbol{y}}|\boldsymbol{f}_{\delta}}^{\dagger}\vec{\mu}_{\vec{\boldsymbol{y}}|\boldsymbol{f}_{\delta}}-2\Re\left[\left\langle \vec{y},\vec{\mu}_{\vec{\boldsymbol{y}}|\boldsymbol{f}_{\delta}}\right\rangle \right]\nonumber \\
 & = & \vec{y}^{\dagger}\vec{y}+\vec{\mu}_{\vec{\boldsymbol{h}}}^{\dagger}\grave{X}^{\dagger}\grave{X}\vec{\mu}_{\vec{\boldsymbol{h}}}-2\Re\left[\left\langle \vec{y},\grave{X}\vec{\mu}_{\vec{\boldsymbol{h}}}\right\rangle \right]\nonumber \\
 & = & \vec{y}^{\dagger}\vec{y}+\vec{\mu}_{\vec{\boldsymbol{h}}}^{\dagger}\grave{S}^{\dagger}\grave{S}\vec{\mu}_{\vec{\boldsymbol{h}}}-2\Re\left[\left\langle \grave{X}^{\dagger}\vec{y},\vec{\mu}_{\vec{\boldsymbol{h}}}\right\rangle \right],\label{eq:yIy}
\end{eqnarray}
and
\begin{eqnarray}
 &  & \left(\vec{y}-\vec{\mu}_{\vec{\boldsymbol{y}}|\boldsymbol{f}_{\delta}}\right)^{\dagger}\grave{X}\left(\Sigma_{\vec{\boldsymbol{h}}}^{-1}+\grave{S}^{\dagger}\grave{S}\right)^{-1}\grave{X}^{\dagger}\left(\vec{y}-\vec{\mu}_{\vec{\boldsymbol{y}}|\boldsymbol{f}_{\delta}}\right)\nonumber \\
 & = & \left(\grave{X}^{\dagger}\vec{y}-\grave{S}^{\dagger}\grave{S}\vec{\mu}_{\vec{\boldsymbol{h}}}\right)^{\dagger}\left(\Sigma_{\vec{\boldsymbol{h}}}^{-1}+\grave{S}^{\dagger}\grave{S}\right)^{-1}\left(\grave{X}^{\dagger}\vec{y}-\grave{S}^{\dagger}\grave{S}\vec{\mu}_{\vec{\boldsymbol{h}}}\right)\nonumber \\
 & = & \left(\grave{X}^{\dagger}\vec{y}\right)^{\dagger}\left(\Sigma_{\vec{\boldsymbol{h}}}^{-1}+\grave{S}^{\dagger}\grave{S}\right)^{-1}\left(\grave{X}^{\dagger}\vec{y}\right)+\nonumber \\
 &  & \left(\grave{S}^{\dagger}\grave{S}\vec{\mu}_{\vec{\boldsymbol{h}}}\right)^{\dagger}\left(\Sigma_{\vec{\boldsymbol{h}}}^{-1}+\grave{S}^{\dagger}\grave{S}\right)^{-1}\left(\grave{S}^{\dagger}\grave{S}\vec{\mu}_{\vec{\boldsymbol{h}}}\right)\nonumber \\
 &  & -2\Re\left[\left(\grave{S}^{\dagger}\grave{S}\vec{\mu}_{\vec{\boldsymbol{h}}}\right)^{\dagger}\left(\Sigma_{\vec{\boldsymbol{h}}}^{-1}+\grave{S}^{\dagger}\grave{S}\right)^{-1}\left(\grave{X}^{\dagger}\vec{y}\right)\right].\label{eq:yXy}
\end{eqnarray}
After discarding terms in \prettyref{eq:yIy} and \prettyref{eq:yXy}
that are not functions of $f_{\delta}$, we obtain \prettyref{eq:gyf}.

\section{\label{sec:Cal-dgdf}Proof of Theorem \ref{thm:dgdf-0} }

We find the derivatives of the three terms in $\frac{\partial g(\vec{y},f_{\delta})}{\partial f_{\delta}}$
of \prettyref{eq:gyf} as follows. The first one is 
\begin{eqnarray}
 &  & \frac{\partial2\Re\left[\left\langle \grave{X}^{\dagger}\vec{y},\vec{b}\right\rangle \right]}{\partial f_{\delta}}\nonumber \\
 & = & 2\Re\left[\left\langle \frac{\partial}{\partial f_{\delta}}\left[S^{\dagger}\left[e^{-j2\pi f_{\delta}(k-1)}y_{r,k}\right]_{k}\right]_{r},\vec{b}\right\rangle \right]\nonumber \\
 & = & 2\Re\left[-j2\pi\left\langle \left[S^{\dagger}\left[(k-1)e^{-j2\pi f_{\delta}(k-1)}y_{r,k}\right]_{k}\right]_{r},\vec{b}\right\rangle \right]\nonumber \\
 & = & 4\pi\Im\left[\left\langle \left[S^{\dagger}\left[(k-1)e^{-j2\pi f_{\delta}(k-1)}y_{r,k}\right]_{k}\right]_{r},\vec{b}\right\rangle \right]\nonumber \\
 & = & 4\pi\Im\left[\sum_{k'=1}^{n-1}e^{-j2\pi f_{\delta}k'}k'\sum_{r,t}s_{t,k'+1}^{*}y_{r,k'+1}b_{r,t}^{*}\right]\nonumber \\
 & = & -4\pi\Im\left[\sum_{k'=1}^{n-1}e^{j2\pi f_{\delta}k'}k'\sum_{r,t}s_{t,k'+1}y_{r,k'+1}^{*}b_{r,t}\right].\label{eq:d1-df}
\end{eqnarray}
The second one is

\begin{eqnarray}
 &  & \frac{\partial\left(\grave{X}^{\dagger}\vec{y}\right)^{\dagger}A\left(\grave{X}^{\dagger}\vec{y}\right)}{\partial f_{\delta}}\nonumber \\
 & = & \frac{\partial\text{Tr}\left(A\left(\grave{X}^{\dagger}\vec{y}\right)\left(\grave{X}^{\dagger}\vec{y}\right)^{\dagger}\right)}{\partial f_{\delta}}\nonumber \\
 & = & \frac{\partial\text{Tr}\left(A\grave{S}^{\dagger}\grave{F}^{\dagger}\vec{y}\vec{y}^{\dagger}\grave{F}\grave{S}\right)}{\partial f_{\delta}}\nonumber \\
 & = & \text{Tr}\left(A\grave{S}^{\dagger}\left[\left[\frac{\partial}{\partial f_{\delta}}e^{j2\pi f_{\delta}(k_{1}-k_{2})}\right.\right.\right.\nonumber \\
 &  & \left.\left.\left.\times y_{r_{2},k_{2}}y_{r_{1},k_{1}}^{*}\right]_{k_{2},k_{1}}\right]_{r_{2},r_{1}}\grave{S}\right)\nonumber \\
 & = & \text{Tr}\left(A\left[S^{\dagger}\left[j2\pi(k_{1}-k_{2})e^{j2\pi f_{\delta}(k_{1}-k_{2})}\right.\right.\right.\nonumber \\
 &  & \left.\left.\left.\times y_{r_{2},k_{2}}y_{r_{1},k_{1}}^{*}\right]_{k_{2},k_{1}}S\right]_{r_{2},r_{1}}\right)\nonumber \\
 & = & j2\pi\sum_{k_{1},k_{2}}e^{j2\pi f_{\delta}(k_{1}-k_{2})}(k_{1}-k_{2})\times\nonumber \\
 &  & \sum_{r_{1},t_{1},r_{2},t_{2}}s_{t_{1},k_{1}}a_{r_{1},t_{1},r_{2},t_{2}}s_{t_{2},k_{2}}^{*}y_{r_{2},k_{2}}y_{r_{1},k_{1}}^{*}\label{eq:full-sum}\\
 & = & -4\pi\Im\left[\sum_{k'=1}^{n-1}e^{j2\pi f_{\delta}k'}k'\sum_{k_{1}=k'+1}^{n}\sum_{r_{1},t_{1},r_{2},t_{2}}\right.\nonumber \\
 &  & \left.a_{r_{1},t_{1},r_{2},t_{2}}s_{t_{1},k_{1}}s_{t_{2},k_{1}-k'}^{*}y_{r_{2},k_{1}-k'}y_{r_{1},k_{1}}^{*}\right],\label{eq:add-lower-triangle-1}
\end{eqnarray}
where \prettyref{eq:add-lower-triangle-1} follows from the observation
that $x-x^{*}=2j\Im[x]$ and that the summand in \prettyref{eq:full-sum}
is anti-symmetric when exchanging $k_{1}$ and $k_{2}$ and thus,
we only need to sum for $k_{1}>k_{2}$, while defining $k'=k_{1}-k_{2}$
and replacing $k_{2}=k_{1}-k'$. The third one is 
\begin{eqnarray}
 &  & -\frac{1}{2}\sigma_{\boldsymbol{f}_{\delta}}^{-2}\frac{\partial}{\partial f_{\delta}}\left|f_{\delta}-\mu_{\boldsymbol{f}_{\delta}}\right|^{2}\nonumber \\
 & = & -\sigma_{\boldsymbol{f}_{\delta}}^{-2}\left(f_{\delta}-\mu_{\boldsymbol{f}_{\delta}}\right).\label{eq:d3-df}
\end{eqnarray}
 Combing eq. (\ref{eq:d1-df}, \ref{eq:add-lower-triangle-1}, \ref{eq:d3-df}),
we obtain eq. (\ref{eq:dg-df-general}, \ref{eq:rtheta-general}).

\section{\label{sec:Proof-CRLB}Proof of Theorem \ref{thm:CRLB} of the Cramer-Rao
Lower Bound \label{sec:Proof-of-Theorem}}

To calculate the BCRLB, calculate
\begin{eqnarray}
 &  & \frac{\partial^{2}\ln\left(f_{\vec{\boldsymbol{y}}|\boldsymbol{f}_{\delta}}(\vec{y}|f_{\delta})f_{\boldsymbol{f}_{\delta}}(f_{\delta})\right)}{\partial f_{\delta}^{2}}\nonumber \\
 & = & \frac{\partial^{2}g(\vec{y},f_{\delta})}{\partial f_{\delta}^{2}}\nonumber \\
 & = & -\frac{\partial}{\partial f_{\delta}}4\pi\Im\left[\sum_{k=1}^{n-1}e^{j2\pi f_{\delta}k}kr_{k}e^{-j\theta_{k}}\right]\nonumber \\
 &  & -\frac{\partial}{\partial f_{\delta}}\sigma_{\boldsymbol{f}_{\delta}}^{-2}\left(f_{\delta}-\mu_{\boldsymbol{f}_{\delta}}\right)\label{eq:used-df-dfd}\\
 & = & -4\pi\Im\left[j2\pi\sum_{k=1}^{n-1}e^{j2\pi f_{\delta}k}k^{2}r_{k}e^{-j\theta_{k}}\right]-\sigma_{\boldsymbol{f}_{\delta}}^{-2}\nonumber \\
 & = & -8\pi^{2}\Re\left[\sum_{k=1}^{n-1}e^{j2\pi f_{\delta}k}k^{2}r_{k}e^{-j\theta_{k}}\right]-\sigma_{\boldsymbol{f}_{\delta}}^{-2},\nonumber 
\end{eqnarray}
where we have used \prettyref{eq:dg-df-general}%
. %

Note that $\text{E}\left[\cdot\right]=\text{E}_{\boldsymbol{f}_{\delta}}\left[\text{E}_{\vec{\boldsymbol{y}}|\boldsymbol{f}_{\delta}}\left[\cdot\right]\right]$.
We calculate 
\begin{eqnarray}
 &  & \text{E}_{\vec{\boldsymbol{y}}|\boldsymbol{f}_{\delta}}\left[\frac{\partial^{2}\ln\left(f_{\vec{\boldsymbol{y}}|\boldsymbol{f}_{\delta}}(\vec{\boldsymbol{y}}|\boldsymbol{f}_{\delta})f_{\boldsymbol{f}_{\delta}}(\boldsymbol{f}_{\delta})\right)}{\partial\boldsymbol{f}_{\delta}^{2}}\right]\nonumber \\
 & = & -8\pi^{2}\Re\left[\sum_{k=1}^{n-1}e^{j2\pi\boldsymbol{f}_{\delta}k}k^{2}\text{E}_{\vec{\boldsymbol{y}}|\boldsymbol{f}_{\delta}}\left[\boldsymbol{r}_{k}e^{-j\boldsymbol{\theta}_{k}}\right]\right]-\sigma_{\boldsymbol{f}_{\delta}}^{-2}\label{eq:Edln}
\end{eqnarray}
 first. Inspecting \prettyref{eq:rtheta-general}, we need to calculate
\begin{eqnarray*}
 &  & \text{E}_{\vec{\boldsymbol{y}}|\boldsymbol{f}_{\delta}}\left[\boldsymbol{y}_{r,k+1}^{*}\right]\\
 & = & \text{E}_{\vec{\boldsymbol{y}}|\boldsymbol{f}_{\delta}}\left[e^{-j2\pi\boldsymbol{f}_{\delta}k}\sum_{t'}s_{t',k+1}^{*}\boldsymbol{h}_{r,t'}^{*}+\boldsymbol{n}_{r,k+1}^{*}\right]\\
 & = & e^{-j2\pi\boldsymbol{f}_{\delta}k}\sum_{t'}s_{t',k+1}^{*}\mu_{\boldsymbol{h}_{r,t'}}^{*}
\end{eqnarray*}
 and 
\begin{eqnarray*}
 &  & \text{E}_{\vec{\boldsymbol{y}}|\boldsymbol{f}_{\delta}}\left[y_{r_{1},k_{1}}^{*}y_{r_{2},k_{2}}\right]\\
 & = & \text{E}_{\vec{\boldsymbol{y}}|\boldsymbol{f}_{\delta}}\left[\left(e^{-j2\pi\boldsymbol{f}_{\delta}(k_{1}-1)}\sum_{t_{1}'}s_{t_{1}',k_{1}}^{*}\boldsymbol{h}_{r_{1},t_{1}'}^{*}+\boldsymbol{n}_{r_{1},k_{1}}^{*}\right)\right.\\
 &  & \left.\left(e^{j2\pi\boldsymbol{f}_{\delta}(k_{2}-1)}\sum_{t_{2}'}s_{t_{2}',k_{2}}\boldsymbol{h}_{r_{2},t_{2}'}+\boldsymbol{n}_{r_{2},k_{2}}\right)\right]\\
 & \stackrel{k_{2}=k_{1}-k}{=} & e^{-j2\pi\boldsymbol{f}_{\delta}k}\sum_{t_{1}'}s_{t_{1}',k_{1}}^{*}\sum_{t_{2}'}s_{t_{2}',k_{1}-k}\text{E}\left[\boldsymbol{h}_{r_{1},t_{1}'}^{*}\boldsymbol{h}_{r_{2},t_{2}'}\right]\\
 &  & +\delta[r_{1}-r_{2}]\delta[k],
\end{eqnarray*}
 where $\text{E}\left[\boldsymbol{h}_{r_{1},t_{1}'}^{*}\boldsymbol{h}_{r_{2},t_{2}'}\right]=c_{\boldsymbol{h}_{r_{1},t_{1}'}\boldsymbol{h}_{r_{2},t_{2}'}}^{*}+\mu_{\boldsymbol{h}_{r_{2},t_{2}'}}\mu_{\boldsymbol{h}_{r_{1},t_{1}'}}^{*}$;
and $c_{\boldsymbol{h}_{r_{1},t_{1}'}\boldsymbol{h}_{r_{2},t_{2}'}}$
is defined in \prettyref{eq:cov-h}. They are used to obtain%
{} 
\begin{eqnarray}
 &  & \text{E}_{\vec{\boldsymbol{y}}|\boldsymbol{f}_{\delta}}\left[\boldsymbol{r}_{k}e^{-j\boldsymbol{\theta}_{k}}\right]\nonumber \\
 & = & e^{-j2\pi\boldsymbol{f}_{\delta}k}\sum_{r,t}\sum_{t'}s_{t,k+1}s_{t',k+1}^{*}\mu_{\boldsymbol{h}_{r,t'}}^{*}b_{r,t}+\nonumber \\
 &  & e^{-j2\pi\boldsymbol{f}_{\delta}k}\sum_{k_{1}=k+1}^{n}\sum_{r_{1},t_{1},r_{2},t_{2}}a_{r_{1},t_{1},r_{2},t_{2}}\times\nonumber \\
 &  & s_{t_{1},k_{1}}s_{t_{2},k_{1}-k}^{*}\sum_{t_{2}'}s_{t_{2}',k_{1}-k}\sum_{t_{1}'}s_{t_{1}',k_{1}}^{*}\times\nonumber \\
 &  & \left(c_{\boldsymbol{h}_{r_{1},t_{1}'},\boldsymbol{h}_{r_{2},t_{2}'}}^{*}+\mu_{\boldsymbol{h}_{r_{2},t_{2}'}}\mu_{\boldsymbol{h}_{r_{1},t_{1}'}}^{*}\right),\ k\ne0.\label{eq:Ertheta}
\end{eqnarray}
Plug \prettyref{eq:Ertheta} into \prettyref{eq:Edln}, we see that
$\text{E}_{\vec{\boldsymbol{y}}|\boldsymbol{f}_{\delta}}\left[\frac{\partial^{2}\ln\left(f_{\vec{\boldsymbol{y}}|\boldsymbol{f}_{\delta}}(\vec{\boldsymbol{y}}|\boldsymbol{f}_{\delta})f_{\boldsymbol{f}_{\delta}}(\boldsymbol{f}_{\delta})\right)}{\partial\boldsymbol{f}_{\delta}^{2}}\right]$
is not a function of $\boldsymbol{f}_{\delta}$. Therefore, 
\begin{eqnarray*}
 &  & \text{E}_{\vec{\boldsymbol{y}}|\boldsymbol{f}_{\delta}}\left[\frac{\partial^{2}\ln\left(f_{\vec{\boldsymbol{y}}|\boldsymbol{f}_{\delta}}(\vec{\boldsymbol{y}}|\boldsymbol{f}_{\delta})f_{\boldsymbol{f}_{\delta}}(\boldsymbol{f}_{\delta})\right)}{\partial\boldsymbol{f}_{\delta}^{2}}\right]\\
 & = & \text{E}_{\boldsymbol{f}_{\delta}}\left[\text{E}_{\vec{\boldsymbol{y}}|\boldsymbol{f}_{\delta}}\left[\frac{\partial^{2}\ln\left(f_{\vec{\boldsymbol{y}}|\boldsymbol{f}_{\delta}}(\vec{\boldsymbol{y}}|\boldsymbol{f}_{\delta})f_{\boldsymbol{f}_{\delta}}(\boldsymbol{f}_{\delta})\right)}{\partial\boldsymbol{f}_{\delta}^{2}}\right]\right],
\end{eqnarray*}
which is plugged into \prettyref{eq:1-E} to obtain $\beta$ of BCRLB
in \prettyref{eq:-E-general}.

One can calculate $\text{E}_{\vec{\boldsymbol{y}}|\{\boldsymbol{f}_{\delta}=f_{\delta}\}}\left[\frac{\partial^{2}\ln\left(f_{\vec{\boldsymbol{y}}|\boldsymbol{f}_{\delta}}(\vec{\boldsymbol{y}}|f_{\delta})\right)}{\partial f_{\delta}^{2}}\right]$
and observe that it is obtained by setting $\sigma_{\boldsymbol{f}_{\delta}}^{-2}=0$.
This gives CRLB.

\begin{itemize}
\item For i.i.d. zero mean channel, \prettyref{eq:-E-zero-channel} is proved
by plugging \prettyref{eq:a-white-channel} into \prettyref{eq:-E-general}
and employing the definition of $\ddot{s}_{k_{1},k_{1}-k}$ in \prettyref{eq:def-sddot}.
\begin{itemize}
\item For periodic pilot, observe that according to \ref{eq:def-sddot},
\begin{eqnarray*}
\left|\ddot{s}_{k_{1},k_{1}-k}\right|^{2} & = & \rho^{2}
\end{eqnarray*}
 is only nonzero for $k=il_{\text{t}}$, $i=1,...,m-1$. Define $k_{1}=(i_{1}-1)l_{\text{t}}+i_{2}$,
$i_{1}=i+1,...,m$, $i_{2}=1,...,l_{\text{t}}$. Then in \ref{eq:-E-zero-channel},
\begin{eqnarray*}
 &  & \sum_{k=1}^{n-1}k^{2}\sum_{k_{1}=k+1}^{n}\left|\ddot{s}_{k_{1},k_{1}-k}\right|^{2}\\
 & = & \sum_{i=1}^{m-1}(il_{\text{t}})^{2}\sum_{i_{1}=i+1}^{m}\sum_{i_{2}=1}^{l_{\text{t}}}\rho^{2}\\
 & = & \rho^{2}l_{\text{t}}^{3}\sum_{i=1}^{m-1}i^{2}(m-i)\\
 & = & \rho^{2}l_{\text{t}}^{3}\left(\frac{m(m-1)m(2m-1)}{6}-\frac{(m-1)^{2}m^{2}}{4}\right)\\
 & = & \rho^{2}l_{\text{t}}^{3}\left(\frac{m^{2}(m^{2}-1)}{12}\right).\\
 & = & \frac{\rho^{2}}{l_{\text{t}}}\left(\frac{n^{2}(n^{2}-l_{\text{t}}^{2})}{12}\right),
\end{eqnarray*}
which produces \ref{eq:-E-zero-channel-periodic}.
\item For time division pilot, observe that according to \ref{eq:def-sddot},
\begin{eqnarray*}
\left|\ddot{s}_{k_{1},k_{1}-k}\right|^{2} & = & \rho^{2}
\end{eqnarray*}
 is only nonzero for $k=i=1,...,m-1$. Define $k_{1}=(i_{2}-1)m+i_{1}$,
$i_{2}=1,...,l_{\text{t}}$, $i_{1}=i+1,...,m$. Then in \ref{eq:-E-zero-channel},
similarly
\begin{eqnarray*}
 &  & \sum_{k=1}^{n-1}k^{2}\sum_{k_{1}=k+1}^{n}\left|\ddot{s}_{k_{1},k_{1}-k}\right|^{2}\\
 & = & \sum_{i=1}^{m-1}(i)^{2}\sum_{i_{1}=i+1}^{m}\sum_{i_{2}=1}^{l_{\text{t}}}\rho^{2}\\
 & = & \frac{\rho^{2}}{l_{\text{t}}^{3}}\left(\frac{n^{2}(n^{2}-l_{\text{t}}^{2})}{12}\right),
\end{eqnarray*}
which produces \ref{eq:-E-zero-channel-periodic}.
\end{itemize}
\end{itemize}

\bibliographystyle{IEEEtran}
\bibliography{Fine_Frequency_Synchronization}

\end{document}